\documentclass[authoryear,12pt]{elsarticle_modified}
\usepackage{graphicx}
\usepackage{amsmath}
\usepackage{amssymb}
\usepackage[linesnumbered,ruled,vlined]{algorithm2e}
\usepackage{natbib}
\usepackage[colorlinks=true, linkcolor=red,
anchorcolor=red, 
citecolor = red]{hyperref}
\usepackage{geometry}
\usepackage{caption}
\usepackage{subcaption}
\usepackage{xcolor}
\usepackage{csquotes}
\usepackage{setspace}
\usepackage{array}
\usepackage{booktabs}

\usepackage{amsthm}   % theorem and proof

\usepackage[most]{tcolorbox}
\usepackage{url}
\usepackage{boxedminipage}
\usepackage{textpos}
\begin{document}

\begin{frontmatter}

\title{\vspace{0.5cm}\textbf{System optimality versus self-organization in metro networks: An optimal transport analysis}}

\author[rvt1,rvt2]{Tianyu Dong\fnref{equal}}
\author[rvt1,rvt2,rvt3]{Jiazu Zhou\fnref{equal}}
\author[rvt2]{Markus Schl\"{a}pfer\corref{cor1}}

\fntext[equal]{Contributed equally to this work.}
\cortext[cor1]{Corresponding author at: Department of Civil Engineering and Engineering Mechanics, Columbia University, New York, NY
10027, USA. \\ \indent \  Email address: m.schlaepfer@columbia.edu}
\address[rvt1]{Future Cities Laboratory Global, Singapore-ETH Centre, Singapore 138602, Singapore}
\address[rvt2]{Department of Civil Engineering and Engineering Mechanics, Columbia University, New York, NY 10027, USA}
\address[rvt3]{Institute of High Performance Computing (IHPC), Agency for Science, Technology and Research (A*STAR), Singapore 138632, Singapore}

\begin{abstract}
\begin{textblock*}{\textwidth}(0cm,-10.6cm)
\begin{center}
\begin{footnotesize}
\setstretch{0.9}
\begin{boxedminipage}{\textwidth}
This is the Accepted Manuscript version of an article published in \emph{Computers, Environment and Urban Systems} (Elsevier), available at: \url{https://doi.org/10.1016/j.compenvurbsys.2026.102479}. Cite as: Dong, T., Zhou, J., \& Schl\"{a}pfer, M. (2026). System optimality versus self-organization in metro networks: An optimal transport analysis. \textit{Computers, Environment and Urban Systems}, \textit{129}, 102479.
\end{boxedminipage}
\end{footnotesize}
\end{center}
\end{textblock*}
Urban transportation network design is typically approached through top-down planning grounded in engineering and economics. Yet cities are complex systems characterized by feedback loops between infrastructure and mobility demand: network structure shapes origin-destination (OD) flows, while OD flows adapt and in turn affect network development. Despite this interdependence, computational tools to benchmark real-world networks against alternative generative principles remain limited. Here, we introduce a data-driven framework to compare three types of metro networks: a self-organized (desire-path) network derived from local rules, a system-optimal network derived from global optimization, and the empirical real-world network. We quantify similarities and discrepancies using geometric comparisons and optimal transport theory. Applying the framework to Singapore's Mass Rapid Transit system, we find that the empirical metro network is substantially closer to the self-organized benchmark than to the system-optimal benchmark, with remaining discrepancies largely explained by geographic constraints. These findings highlight that system-wide optimality alone may be inadequate for guiding practical interventions, motivating planning approaches that explicitly incorporate local service needs. The framework is transferable across regions and can diagnose design-use misalignments to support adaptive infrastructure planning.
\end{abstract}


\end{frontmatter}

\section{Introduction}

Urban transportation networks form an important backbone of our cities, enabling the spatial interactions that underpin urban social and economic activity \citep{batty2013new, bettencourt2021introduction}. Over time, these networks have been shaped by societal needs and technological advances, leading to substantial heterogeneity across regions that reflects geography, economic development, and culture \citep{batty2008size, diao2021impacts}. This diversity underscores the complexity of designing transportation networks that remain effective across heterogeneous and evolving demands.

Most transportation network design research is grounded in the pursuit of system-wide optimal performance. In transportation engineering, this often entails identifying network configurations that minimize travel distance or time and/or maximize accessibility, subject to budget and feasibility constraints \citep{luathep2011global, tong2015transportation,FARBER201730,kujala2018travel}. In economics and operations research, network interventions are typically evaluated through cost-benefit and welfare lenses, such as road pricing and public transport policies designed to internalize congestion externalities \citep{beaudoin2015public, farahani2013review}. More recently, network design has increasingly been formulated as a multi-objective optimization problem that simultaneously considers cost, travel time, and emissions \citep{ahern2022approximate, liu2020pareto, hosseininasab2018multi, wu2024multi}. Many approaches adopt bi-level formulations, where the lower level represents users' route choices through traffic assignment, and the upper level optimizes network design decisions to achieve a system-level objective~\citep{pinto2020joint}.

Beyond normative optimization, social and behavioral research emphasizes that transportation outcomes depend on adaptive travel decisions (e.g., route choice and departure time) that are difficult to predict and often require costly surveys and large datasets \citep{mcfaddenc2007behavioral, song2010limits}. When formal infrastructure does not meet local needs, bottom-up adaptations can emerge. For instance, pedestrians may create \enquote*{desire paths}, which are informal routes that reveal misalignments between planned design and actual use \citep{helbing1997modelling,foster2019detroit}.

Most existing approaches to transportation network design, across engineering, economics, and related social-science perspectives, implicitly adopt a top-down paradigm in which infrastructure decisions are formulated centrally and implemented through planning institutions \citep{batty2008size}. At the same time, a growing body of research conceptualizes cities as prototypical complex systems \citep{batty2016complexity, jiang2017activity}, where macroscopic structure emerges from the interactions of many heterogeneous agents. Transportation systems exemplify this complexity: network structure shapes OD flows, while mobility demand feeds back into network development. This coupled feedback implies that network design cannot be understood solely as a static optimization problem, but should also be studied as an adaptive process driven by bottom-up dynamics. Despite this perspective, computational frameworks for bottom-up transportation network planning remain relatively scarce, and many models still rely on post-optimization assumptions such as user equilibrium, potentially overlooking heterogeneity in preferences \citep{han2015elastic}. A key open question is therefore whether real-world metro networks more closely resemble structures implied by system-optimal design or by self-organizing travel behavior.

In this work, we introduce a data-driven framework to quantify similarities and discrepancies among three metro network structures: (i) a self-organized network, (ii) a system-optimal network, and (iii) the empirical real-world network. Our self-organized (local rule-based) network, which we also refer to as the \enquote{desire-path network} (see Section 2), is a minimal model that reflects two generic tendencies of human navigation: a preference for short travel times and the emergence of shared high-use corridors that enable faster, higher-capacity modes and increased service frequency, reducing generalized cost per traveler \citep{schlapfer2021universal}. The system-optimal network follows \citet{bontorin2024emergence} and minimizes aggregate travel time over all trips subject to an infrastructure cost constraint, where link width (and thus speed) serves as the design variable. Both model networks are derived (\enquote{reverse-engineered}) from observed origin-destination (OD) patterns. These patterns are shaped by the existing infrastructure, yet do not explicitly encode network structure in our modeling framework. To compare the resulting networks, we apply geometric metrics and optimal transport theory \citep{leite2024simopt}.

The contribution of this work is threefold. First, we provide a computational framework to benchmark empirical metro networks against alternative generative principles (self-organization and system optimality) using OD flows derived from anonymized mobile phone data, illustrated for Singapore's Mass Rapid Transit (MRT) system. Second, we quantify discrepancies in network topology, highlighting structural differences that are not captured by conventional performance indicators alone. Third, the framework is general and can be applied to other infrastructure networks and emerging mobility systems, enabling comparative assessments of design-use misalignments across different urban contexts.

The remainder of this paper is structured as follows. We first define the self-organized network model based on shortest-path navigation and trajectory bundling. We then introduce the system-optimal network design model and describe our optimal transport-based approach for quantifying similarities and discrepancies between the self-organized, system-optimal, and empirical networks. Next, we apply the framework to Singapore's Mass Rapid Transit (MRT) system using anonymized mobile phone GPS data. Finally, we summarize the main findings, discuss limitations, and outline implications for urban transportation network planning.

\section{Self-organized transportation network}

Transportation systems comprise multiple layers, including rail, roads, cycling infrastructure, and pedestrian networks \citep{alessandretti2023, pappalardo2023future}. Most research focuses on the existing physical infrastructure, yet actual travel behavior does not always follow predetermined structures. A canonical example is the emergence of pedestrian \enquote*{desire paths}, where people form informal shortcuts across open space despite the presence of designed walkways \citep{helbing1997modelling,lidwell2010universal}. Such self-organized routing patterns have primarily been studied for walking and cycling \citep{keller2013will} and in informal paratransit systems with ad-hoc routes \citep{mittal2024efficient}. Here, we extend the desire-path perspective to metro network structures by \enquote{reverse-engineering} a desire-path network directly from OD flows. We refer to the resulting benchmark as the \enquote{desire-path network} throughout, and use \enquote{self-organized} in the broader sense of structural outcomes consistent with bottom-up tendencies. In doing so, we abstract from detailed engineering and right-of-way constraints to isolate the bottom-up mechanisms that shape route formation.

Indeed, any transportation network may exhibit a degree of self-organization as a complex system in which repeated individual travel decisions give rise to structured, system-level patterns \citep{batty2016complexity, gerrits2024towards, levinson2006self}. Empirical regularities of aggregate mobility, such as gravity-type models, the visitation law, and radiation models, capture aspects of how populations distribute trips across space \citep{erlander1990gravity, schlapfer2021universal, simini2012universal}. At the individual level, travel choices are often modeled as utility-based decisions, whereas a first-order behavioral tendency is the preference for routes that reduce generalized cost (e.g., travel distance, travel time, energy) \citep{ben1985discrete, hasnine2018dynamics, mwale2022factors}. Importantly, efficient routes often combine individualized segments with shared high-use corridors: when many travelers concentrate along similar routes, flows can be consolidated (\enquote*{bundled}) into higher-capacity services (e.g., buses or rail) and higher service frequency, yielding economies of scale and reduced generalized cost per traveler \citep{vuchic2017urban}. In the following, we operationalize these principles through a network construction model.

\subsubsection*{Shortest path (Euclidean baseline)} Let the OD demand be represented by a set of $N$ OD pairs, $\{ (\mathbf{o}_i, \mathbf{d}_i) \}_{i=1}^{N}$, where \(\mathbf{o}_i\) and \(\mathbf{d}_i\) are the coordinates of the origin and destination denoted by longitude and latitude, \((x_i, y_i)\). For each OD pair, we denote the corresponding route by $\mathbf{r}_i$, represented as an ordered sequence of locations connecting $\mathbf{o}_i$ to $\mathbf{d}_i$. The generalized travel cost, such as time, monetary cost, or energy expenditure, is assumed to increase linearly with the total distance of \(\mathbf{r}_i\).  Under this assumption, the optimal route for OD pair $i$ is the straight-line (Euclidean) connection, \(\mathbf{r}_i=\mathbf{o}_i \rightarrow \mathbf{d}_i\) with distance \(||\mathbf{o}_i-\mathbf{d}_i||_2\).

 \subsubsection*{Trajectory bundling} 
 To model the formation of shared high-use corridors, we apply the partition-and-group trajectory clustering framework developed by \citet{lee2007trajectory}. At initial, the route choice corresponds to the shortest path, \(\mathbf{r}_i=\mathbf{o}_i \rightarrow \mathbf{d}_i\). The whole route is then partitioned by sampling distance \(S\) into \(\lceil\frac{||o_i-d_i||_2}{S}\rceil\)
 segments and thus discretized as \(\mathbf{r}_i=\mathbf{o}_i \rightarrow \mathbf{m}_i^1 \rightarrow \mathbf{m}_i^2 ... \rightarrow \mathbf{m}_i^{J_i} \rightarrow \mathbf{d}_i\), where \(\mathbf{m}_i^j\) are the intermediate points and $J_i = \lceil\frac{||o_i-d_i||_2}{S}\rceil-1$. The resulting number of points across the \(N\) OD pairs is $P=\sum_{i=1}^{N} J_i$.

The 2-dimensional weighted kernel density estimation (KDE), which is an extension of the standard kernel density estimation, allows individual data points to contribute differently to the estimated probability density function (PDF) based on their frequency. The formula for a two-dimensional weighted KDE is given by: 
\begin{equation}
\begin{aligned}
\hat{f}(x, y) &= \frac{1}{P}\sum_{i=1}^{N}\sum_{j=1}^{J_i} w_i K_h(x - x_i^j, y - y_i^j) \\ &= \frac{1}{Ph^2}\sum_{i=1}^{N} \sum_{j=1}^{J_i} w_i K\left(\frac{x - x_i^j}{h}, \frac{y - y_i^j}{h}\right),
\end{aligned}
\end{equation}
where \(\hat{f}(x, y)\) represents the probability density function of location \((x,y)\), and \(K\) is a two-dimensional kernel function, which is typically constructed as the product of two one-dimensional kernels. The corresponding scaled kernel \(K_h\) is commonly defined as \( K_h(x, y) = \frac{1}{h^2}K\left(\frac{x}{h}, \frac{y}{h}\right) \), where \(h\) is the bandwidth parameter controlling the spatial smoothing (i.e., the width of the kernel). The weights $w_i$ account for the frequency (or intensity) of OD pair $i$ in the dataset, ensuring that high-demand OD pairs contribute proportionally more to the density estimate.

To capture the emergence of shared high-use corridors, we apply a point-advection step in which intermediate points are iteratively displaced toward regions of higher trajectory density. Similar approaches have been used to visualize urban arterial road networks \citep{holten2009force}. For each intermediate point \( \mathbf{m}(x, y) \) with a density estimate \( \hat{f}(x, y) \), the new position \( (x', y') \) of the point after moving a step size \(\eta\) along the gradient of the density estimate can be expressed as:
\begin{subequations}\label{eq:advection}
\begin{align}
x' &= x + \eta \,\frac{\partial \hat{f}}{\partial x}(x,y), \label{eq:advection_x}\\
y' &= y + \eta \,\frac{\partial \hat{f}}{\partial y}(x,y), \label{eq:advection_y}
\end{align}
\end{subequations}
where \( \frac{\partial \hat{f}}{\partial x}(x, y) \) and \( \frac{\partial \hat{f}}{\partial y}(x, y) \) are the gradients of \( \hat{f} \) at \( (x, y) \) in the \( x \) and \( y \) directions, respectively. Consequently, the reconstructed route for each OD demand is updated to  $\mathbf{r}_i'= \mathbf{o}_i \rightarrow \mathbf{m}{_i^1}' \rightarrow \mathbf{m}{_i^2}' ... \rightarrow \mathbf{m}{_i^{J_i}}' \rightarrow \mathbf{d}_i$.

The KDE calculation is then performed again using the reconstructed route, and the new probability density function is \(\hat{f'}(x, y)\). The total variation distance (TVD) between \(\hat{f}(x, y)\) and \(\hat{f'}(x, y)\) is calculated as,
\begin{align}
TVD(\hat{F}, \hat{F}') = \frac{1}{2} \int \int |\hat{f}(x, y) - \hat{f'}(x, y)| \, dx\, dy.
\end{align}

If TVD is larger than a certain threshold \(\epsilon\), it means the current routes do not achieve the converged state, and the points should be advected again. The intermediate points advection should stop when  TVD is smaller than the specific threshold, thereby avoiding overconvergence. Algorithm 1 shows the detailed procedure and iteration process for the trajectory clustering.

\begin{algorithm}[b!]
\caption{Trajectory Bundling}
\label{bundling_algorithm}

\KwIn{Origin--Destination demands $\{O_i, D_i\}$}
\KwOut{Bundled trajectories $\{P_i\}$}

\ForEach{$(O_i, D_i)$}{
    Connect $O_i$ to $D_i$ to form initial path $P_i$\tcp*[r]{Shortest path}
}

\While{not converged}{
    \ForEach{$P_i$}{
        Sample $P_i$ into points ${Point}_i$ at fixed interval $l$
    }

    \ForEach{$\mathrm{Point}_i$}{
        Perform weighted kernel density estimation (KDE)
    }

    \ForEach{$\mathrm{Point}_i$}{
        Move non-terminal points along the weighted KDE gradient\;
        Apply Laplacian smoothing to obtain ${ReconstructedPath}_i$
    }

    Compute total variation distance (TVD) between ${ReconstructedPath}_i$ and $P_i$\;

    \If{TVD $< \varepsilon$}{
        \textbf{break}\tcp*[r]{Convergence achieved}
    }

    Update $P_i \leftarrow \mathrm{ReconstructedPath}_i$\;
}

\end{algorithm}

\subsubsection*{Network extraction}

After the trajectory bundling, we extract a sparse desire-path network from the aggregated set of routes by constructing a weighted graph $G_h=(V,E)$, where \(V\) denotes the set of nodes and \(E\) represents the set of edges connecting these nodes. To that end, we identify candidate network nodes as local maxima (high-density locations) in the kernel density estimate $f(x,y)$. Specifically, the set of nodes is defined as
\begin{align}
V=\{\mathbf{c}=(x,y)\mid f(x,y)>T_d\},
\end{align}
where $T_d$ is a density threshold used to filter out sparsely visited locations. To construct the edges, each intermediate point $\mathbf{m}{_i^j}'$ is assigned to its nearest node
\begin{align}
\mathbf{c}(\mathbf{m}{_i^j}')=\arg\min_{\mathbf{c}_k\in V}\, d\!\left(\mathbf{m}{_i^j}',\mathbf{c}_k\right),
\end{align}
and each route is mapped to a sequence of nodes,
\begin{align}
\mathbf{r}''_i=\mathbf{o}_i \rightarrow \mathbf{c}(\mathbf{m}{_i^1}') \rightarrow \cdots \rightarrow \mathbf{c}(\mathbf{m}{_i^{J_i}}') \rightarrow \mathbf{d}_i .
\end{align}

For any consecutive pair of nodes $(\mathbf{c}_k,\mathbf{c}_\ell)$ appearing in $\mathbf{r}''_i$, we add a directed edge $e_{k\ell}\in E$. The edge weight $w_{k\ell}$ is defined as the number of occurrences of the transition $\mathbf{c}_k\rightarrow \mathbf{c}_\ell$ across all compressed routes $\{\mathbf{r}''_i\}_{i=1}^N$. The resulting weighted graph $G_h$ constitutes the extracted desire-path network.

\section{System-optimal transportation network}
\label{sec:system-optimal}

In this section, we construct a system-optimal benchmark network following the framework of \citet{bontorin2024emergence}. Unlike many studies that optimize modifications to an existing network (e.g., capacity expansion or pricing interventions), we derive an optimized network structure directly from origin--destination (OD) demand, without conditioning on the current infrastructure topology. This design benchmark provides a reference for how a network could be configured to minimize aggregate travel time under an infrastructure cost constraint. We emphasize that OD patterns are, in practice, influenced by the deployed transportation system; nevertheless, using observed OD flows allows us to compare idealized design principles against real-world structure within a consistent demand setting.
    \begin{figure}[t!]
       \centering
       \includegraphics[scale=0.4]{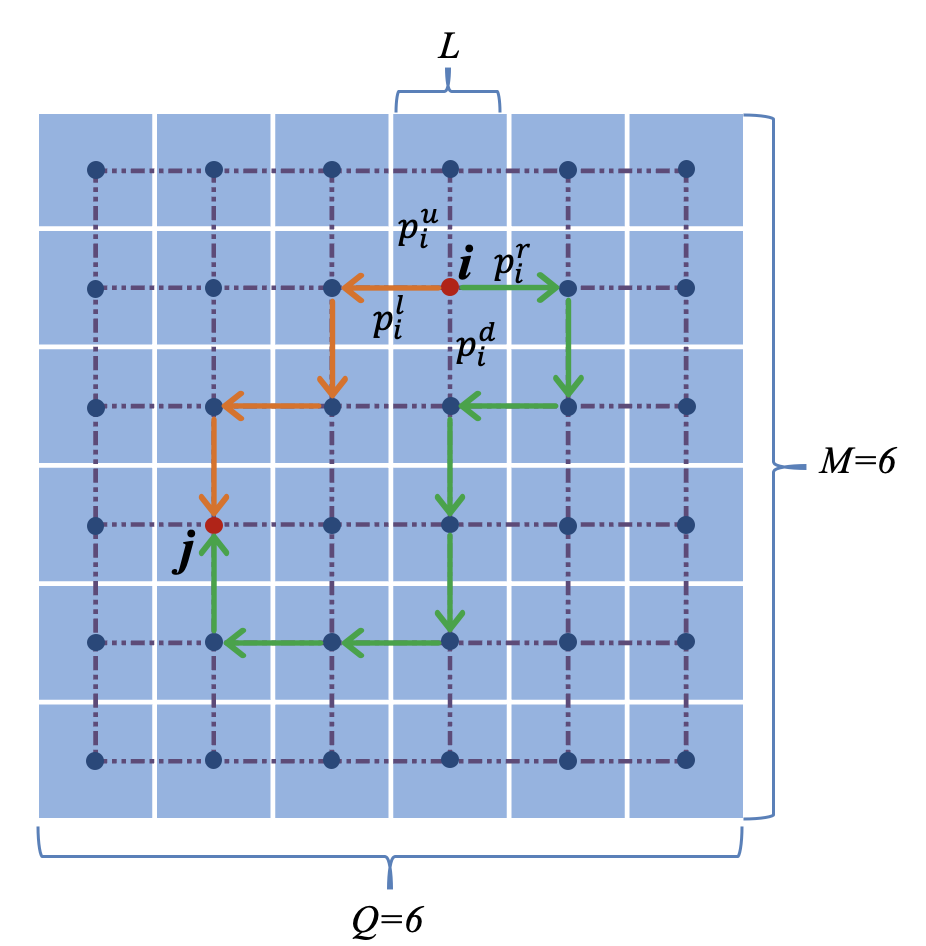}
       \caption{Schematic illustration of the spatial grid representation and the definition of feasible grid paths between OD cells.}
       \label{Fig: example of a spatial gridded network}
    \end{figure}
To reduce computational complexity, we represent the study area on a regular grid following \citet{bontorin2024emergence}. As illustrated in Fig.~\ref{Fig: example of a spatial gridded network}, the area is discretized into an $M \times Q$ lattice with square cells of side length $L$. Each grid cell is indexed by $g_i$ and represented by the coordinates $(x_i,y_i)$ of its cell center. Adjacent cells are connected by edges of length $L$ (4-neighborhood connectivity), yielding a total of $2MQ - M - Q$ undirected edges. We assign a nonnegative weight $w_k$ to each edge $e_k$, interpreted as the travel speed along that link, subject to bounds $0 \le w_k \le w_{\max}$. The travel time on edge $e_k$ is then $t_k={L} / {w_k}$.

Given an OD dataset $\{(\mathbf{o}_n,\mathbf{d}_n)\}_{n=1}^{N}$, we map origins and destinations to their corresponding grid cells, producing OD pairs $\{(g_i,g_j)\}$ with associated demand $OD_{ij}$. For any OD pair $(g_i,g_j)$, let $P_{ij}$ denote the set of feasible grid paths connecting the two cells. Assuming travelers choose minimum-travel-time paths, the shortest travel time between $g_i$ and $g_j$ is computed as
\begin{equation}
T_{ij}=\min_{p_{ij}\in P_{ij}} \sum_{e_k \in p_{ij}} \frac{L}{w_k}.
\label{eq:shortesttime_grid}
\end{equation}

The system-optimal design problem seeks edge weights $\{w_k\}$ that minimize the aggregate travel time across all OD flows, constrained by a cost budget:
\begin{equation}
\min_{\{w_k\}} \,\, \sum_{i}\sum_{j\neq i} OD_{ij}\, T_{ij},
\qquad \text{s.t. } Cost(\{w_k\}) \le C_{\max},
\label{eq:system_opt_objective}
\end{equation}
where $C_{\max}$ denotes the budget threshold. Following \citet{bontorin2024emergence}, total infrastructure cost increases with link speeds and can be expressed as a function of $\{w_k\}$,
\begin{equation}
Cost(\{w_k\})=\sum_{k=1}^{2MQ-M-Q} w_k.
\label{eq:cost_constraint}
\end{equation}
We solve this non-linear optimization problem using simulated annealing coupled with repeated shortest-path computations.
    
\section{Comparison of transportation networks} 
We compare network structures using two complementary perspectives. First, we quantify geometric similarity using the Hausdorff distance \citep{huttenlocher1993comparing, taha2015efficient}, following the approach in \citet{meyur2022ensembles}. Second, we employ optimal transport (OT) to obtain a global discrepancy measure between network representations interpreted as distributions over space \citep{santambrogio2015optimal, levy2018notions}.

\subsection{Geometric comparison via Hausdorff distance}
Let $A$ and $B$ denote two sets of sampled points (e.g., network nodes or polyline samples) in a metric space $(X,d)$. The Hausdorff distance between $A$ and $B$ is defined as
\begin{equation}
d_H(A,B)=\max\left\{
\sup_{a\in A}\inf_{b\in B} d(a,b),\;
\sup_{b\in B}\inf_{a\in A} d(b,a)
\right\}.
\label{eq:hausdorff}
\end{equation}
Intuitively, $d_H(A,B)$ measures the maximum distance from a point in one set to its nearest neighbor in the other set.

\subsection{Global comparison via optimal transport}

Geometric summaries alone may miss global differences in how network ``mass'' (importance) is distributed across space. OT provides a principled way to quantify the minimal cost required to transform one distribution into another \citep{santambrogio2015optimal, levy2018notions}. Recent work has demonstrated the usefulness of OT for quantifying similarity and economies of scale in urban transportation networks and for deriving OT-based infrastructure benchmarks \citep{leite2024simopt}. Let the real-world network be $G_r=(V_r,E_r)$ and a comparison network be $G_h=(V_h,E_h)$, where each node is associated with spatial coordinates. We denote node coordinates by
\[
\mathbf{x}^{(r)}_i \in \mathbb{R}^2,\quad i=1,\ldots,|V_r|,
\qquad
\mathbf{x}^{(h)}_j \in \mathbb{R}^2,\quad j=1,\ldots,|V_h|.
\]
We define the OT ground cost between nodes as the Euclidean distance
\begin{equation}
\Pi_{ij}=\|\mathbf{x}^{(r)}_i-\mathbf{x}^{(h)}_j\|_2.
\label{eq:ot_cost}
\end{equation}

\paragraph{Node weights (network mass)}
To convert each network into a probability distribution over its nodes, we assign node weights based on weighted degree (strength). Let $\Gamma_r$ and $\Gamma_h$ denote the weighted adjacency matrices of $G_r$ and $G_h$ (with $\Gamma(i,j)$ equal to the weight of the edge from node $i$ to node $j$, and $0$ if no edge exists). The node strength is
\begin{equation}
s^{(r)}_i=\sum_{j=1}^{|V_r|}\Gamma_r(i,j),
\qquad
s^{(h)}_j=\sum_{\ell=1}^{|V_h|}\Gamma_h(j,\ell).
\label{eq:node_strength}
\end{equation}
We normalize strengths into probability distributions
\begin{equation}
p^{(r)}_i=\frac{s^{(r)}_i}{\sum_{u=1}^{|V_r|} s^{(r)}_u},
\qquad
p^{(h)}_j=\frac{s^{(h)}_j}{\sum_{v=1}^{|V_h|} s^{(h)}_v}.
\label{eq:node_probs}
\end{equation}

Note that edge weights underlying networks compared via OT may carry different physical meanings, for instance, flow-based (passenger volumes) versus capacity-based (travel speed). However, the OT comparison does not directly operate on edge weights but on the normalized node-strength distributions defined above. Because both distributions are normalized to sum to one, the OT distance compares dimensionless probability measures, and the physical units of the original edge weights do not enter the transport cost calculation. In this representation, node strength serves as a generic proxy for the relative importance of locations in a network, regardless of whether that importance arises from demand (flows) or supply (capacity).

\paragraph{Optimal transport problem}
We compute the OT distance by solving the discrete Kantorovich formulation
\begin{equation}
\min_{\gamma \in \mathbb{R}_+^{|V_r|\times |V_h|}}
\sum_{i=1}^{|V_r|}\sum_{j=1}^{|V_h|}\Pi_{ij}\,\gamma_{ij}
\label{eq:ot_objective}
\end{equation}
subject to the marginal constraints
\begin{equation}
\sum_{j=1}^{|V_h|}\gamma_{ij}=p^{(r)}_i \quad \forall i,
\qquad
\sum_{i=1}^{|V_r|}\gamma_{ij}=p^{(h)}_j \quad \forall j,
\qquad
\gamma_{ij}\ge 0 \quad \forall i,j .
\label{eq:ot_constraints}
\end{equation}
The optimal value provides a global discrepancy measure between networks, interpretable as the minimal ``effort'' required to transport network mass from $G_r$ to match $G_h$. This linear program can be solved using standard OT solvers.

\section{Case study application}

We apply the proposed framework to Singapore as a case study using two complementary datasets. We use an individual-level, GPS-based mobile phone dataset from CityData.ai \citep{CityDataAI} to infer general urban OD demand patterns across all modes (walking, cycling, driving, bus, and rail), and an aggregated OD dataset for Mass Rapid Transit (MRT) trips from Singapore's Land Transport Authority (LTA) DataMall \citep{LTADataMall} to represent the empirical metro system. These datasets allow us to construct self-organized (desire-path) and system-optimal benchmark networks and to compare them against the real-world MRT network. We deliberately use multi-modal OD data for the benchmarks, rather than MRT-specific flows. The latter would anchor the benchmarks to the current MRT topology and provide no independent reference for evaluating the network's structure with respect to broader urban mobility needs.

\subsection{Data description}

The mobile phone dataset consists of GPS traces from 894,327 anonymized users in Singapore, collected between September 1 and September 30, 2019. Because sampling rates vary across users, we restrict the analysis to active users, defined as individuals appearing on at least 5 days and contributing more than 120 records over the month \citep{10508873}. This yields 244,515 active users and 218,248,627 GPS records.

\begin{figure}[!b]
    \centering
    \includegraphics[width=0.61\columnwidth]{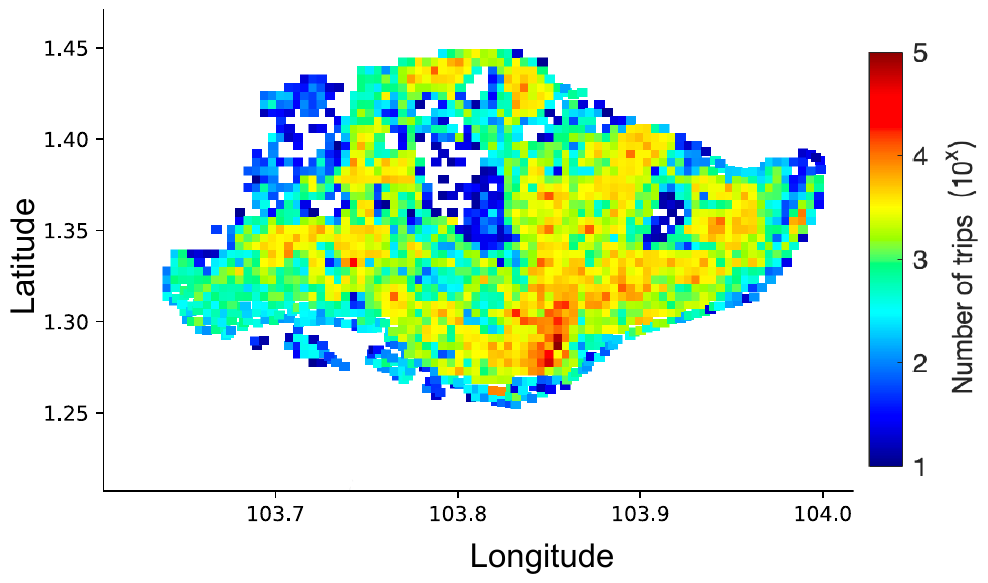}
    \caption{Spatial heatmap of trip volumes in Singapore inferred from GPS traces. Colors indicate the number of trips generated at each origin grid cell.}
    \label{od_from_gps}
\end{figure}

To infer OD demand, Singapore is discretized into $500\,\mathrm{m}\times500\,\mathrm{m}$ grid cells. For each user, we identify stops as locations where the user remains within the same grid cell for at least 30 minutes. The resulting sequence of stops is used to construct OD trips, yielding 3,932,249 trips across 1,005,396 unique OD pairs. To streamline the analysis and focus on trips that are relevant at the metro-network scale, we filter OD pairs shorter than 1 km and OD pairs observed fewer than three times during the study period. After filtering, the dataset contains 2,762,943 OD trips across 225,392 unique OD pairs. Figure~\ref{od_from_gps} illustrates the spatial distribution of trip volumes generated at each origin location.

\begin{figure}[!htb]
    \centering
    \includegraphics[scale=0.24]{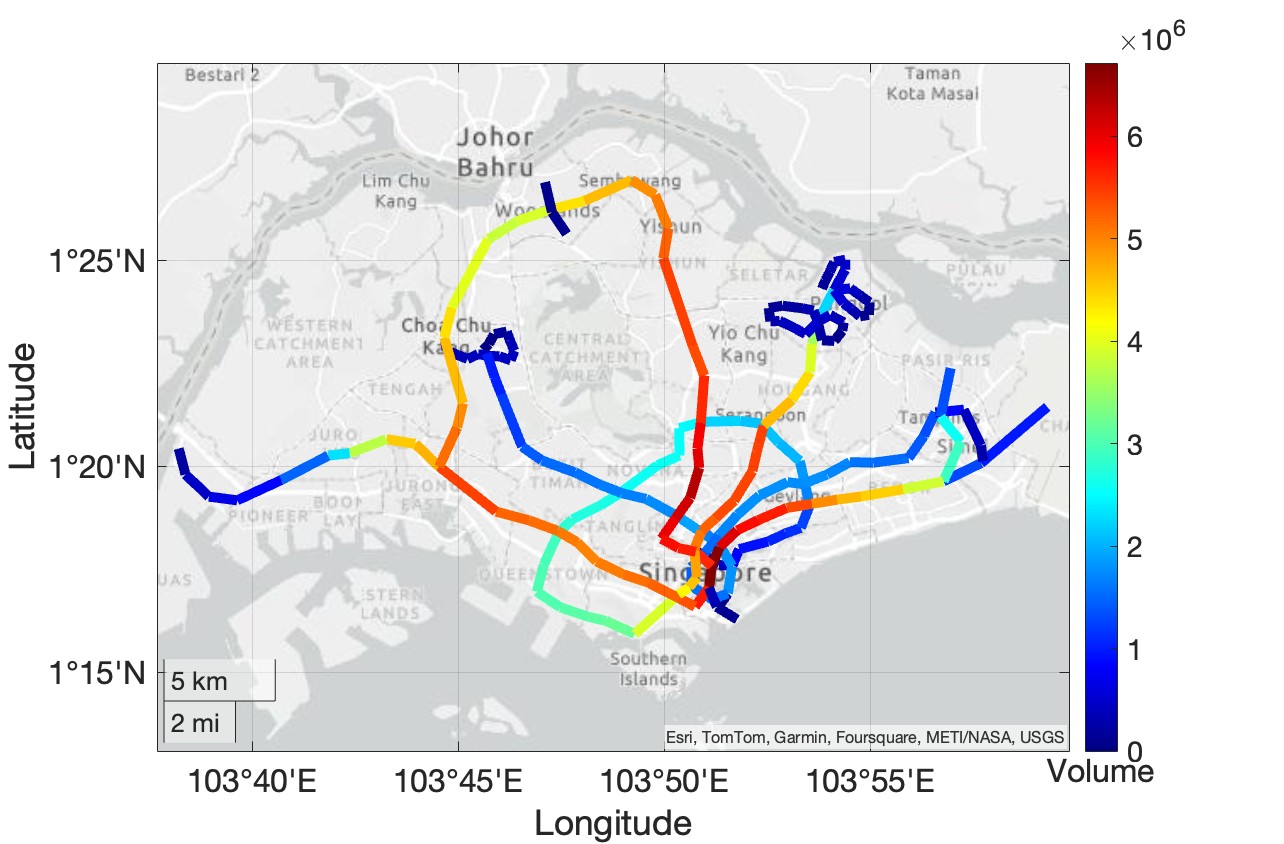}
    \caption{Singapore's Mass Rapid Transit (MRT) network (including its two associated Light Rail Transit lines). Link colors indicate passenger trip volumes in December 2023.}
    \label{sg_mrt}
\end{figure}

The LTA DataMall OD dataset reports the number of MRT passenger trips between origin and destination stations, disaggregated by hour and by weekday/weekend. For our analysis, we use December 2023 as the study period, which contains 76,610,749 MRT trips across 20,223 unique origin-destination station pairs. To map each OD pair onto the network, we assign routes on the empirical MRT graph following the shortest-path principle with a minimal number of interchanges. Figure~\ref{sg_mrt} visualizes the Singapore MRT network together with passenger trip volumes.

Note that there is a time gap between the MRT data (December 2023) and the GPS-derived mobility data (September 2019). The validity of comparing the resulting networks relies primarily on the stability of the \textit{spatial structure} of the mobility demand rather than exact agreement in travel volumes. While total mobility levels fluctuated during and after COVID-19, available evidence suggests that Singapore’s dominant mobility corridors remained largely stable over this period. In particular, MRT ridership levels in 2023 recovered to pre-pandemic levels (details in Appendix A), and no major relocation of residential or employment centers has occurred \citep{urapreviousplan}. Singapore’s geographically constrained and compact urban form further limits large-scale shifts in spatial interaction patterns. As a result, the large-scale structures that underpin the network comparison remain robust. Temporally matched datasets would be desirable but are not available to the authors; the current combination therefore represents a practical approximation.

\subsection{Desire-path transportation network in Singapore}

\begin{figure}[!t]
    \centering
    \includegraphics[width=0.90\linewidth]{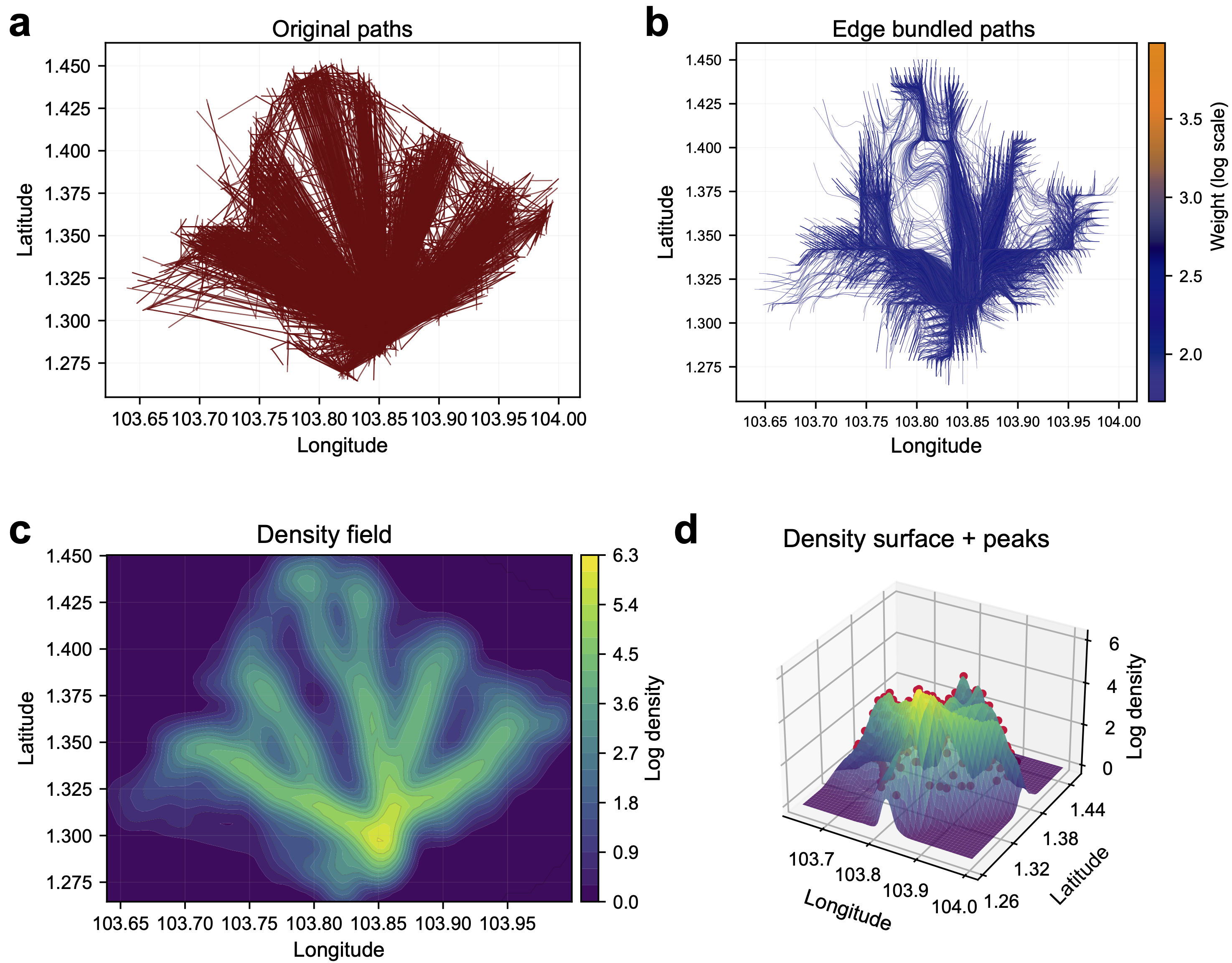}
    \caption{Illustration of the desire-path network construction in Singapore. a) Original shortest paths inferred from the OD data; b) Bundled paths after the point-advection process; c) Density field of bundled trajectories; d) 3D surface of the trajectory density field with detected network nodes.}
    \label{fig:desire_path_construction}
\end{figure}

\begin{figure}[t!]
    \centering
    \includegraphics[width=0.75\textwidth,trim={0 0 0 32},clip]{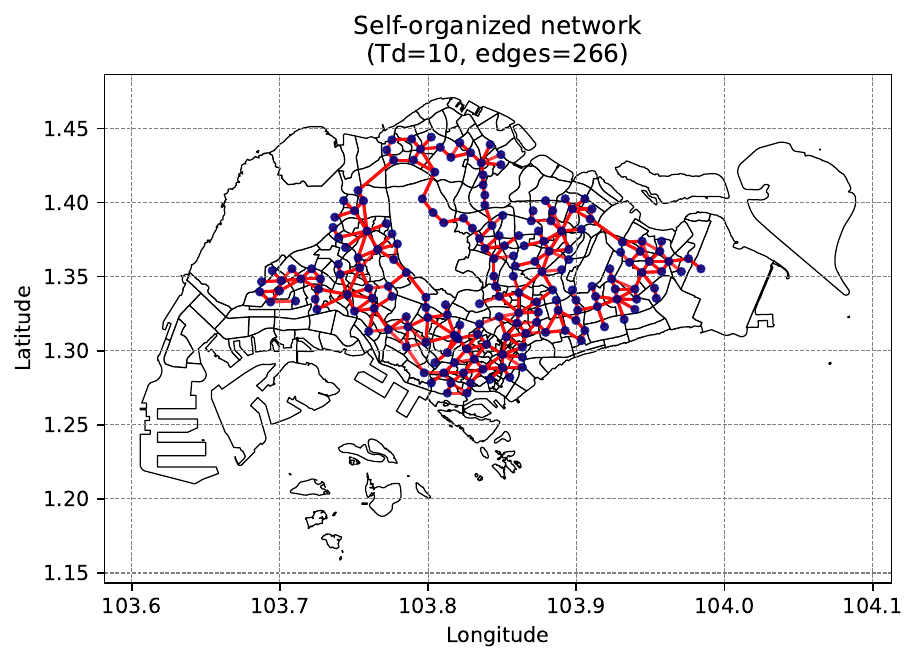}
    \caption{Extracted desire-path network in Singapore.}
    \label{self}
\end{figure}

We apply the trajectory bundling procedure introduced above to model the emergence of shared high-use corridors through route reinforcement. As shown in Fig.~\ref{fig:desire_path_construction}, OD paths are approximated as straight-line connections and bundled using the point-advection process. Prominent peaks in the resulting trajectory density field are identified as network nodes, representing major mobility centers. The KDE bandwidth is set to $h = 500$\,m, matching the spatial resolution of the OD data. We identify desire-path network nodes as local maxima of the KDE density surface that exceed a threshold $T_d = 10\%$ of the maximum density. Sensitivity analyses for both $h$ (varied from 250\,m to 3000\,m) and $T_d$ (varied from 5\% to 20\%) are reported in Appendix~B and confirm that the main conclusions are robust to these choices. The resulting \enquote{desire-path} network for Singapore is shown in Fig.~\ref{self}.

\subsection{System-optimal transportation network in Singapore}

\begin{figure}[t!]
    \centering
    \includegraphics[width=0.75\textwidth, trim={0 0 432 20},clip]{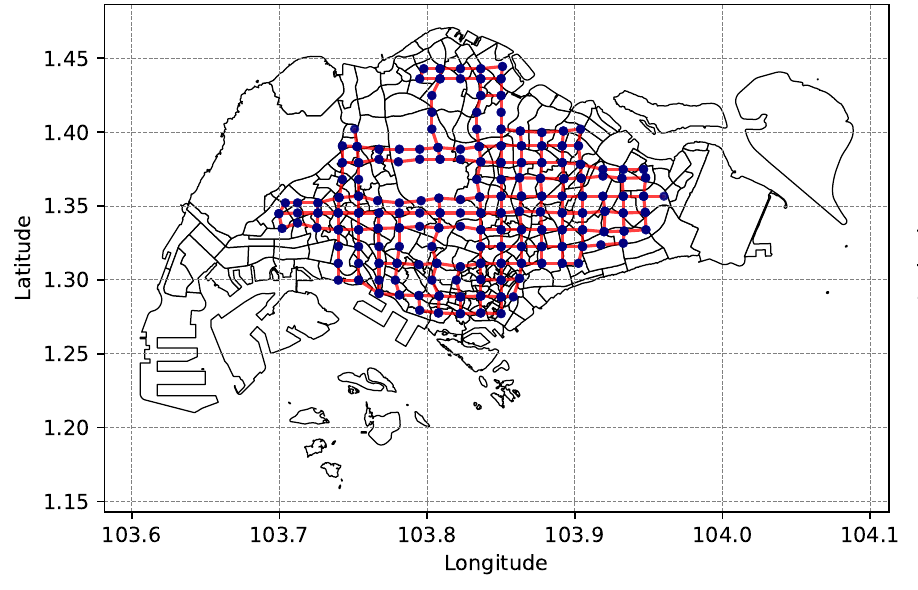}
    \caption{System-optimal benchmark network in Singapore.}
    \label{figurelabel_1}
\end{figure}

To construct the system-optimal benchmark network for Singapore, we discretize the study area into regular grid cells with side length $L$. To minimize artifacts from grid discretization, we use $L = 500$\,m, matching the grid cell size used to process the raw GPS data. We further apply a node-merging post-processing step that merges connected nodes whose inter-node distance falls below 1.3 km, corresponding to the average inter-station spacing of the MRT network, ensuring comparable spatial granularity across all networks. We assess the sensitivity of the system-optimal network to grid resolution by testing $L \in \{500, 750, 1000\}$\,m. The resulting benchmark networks exhibit consistent structural patterns (details in Appendix~B). We then solve the system-optimal design problem described in Section~\ref{sec:system-optimal} to obtain the network that minimizes aggregate travel time under the infrastructure budget constraint. For comparability, parameters in both benchmark models (desire-path and system-optimal) were chosen to yield metro-scale networks with similar spatial footprint and sparsity. The resulting system-optimal network is shown in Fig.~\ref{figurelabel_1}.

\subsection{Comparison of benchmark networks with the empirical MRT network}

\begin{figure}[t!]
   \centering   
   \includegraphics[width=0.95\textwidth]{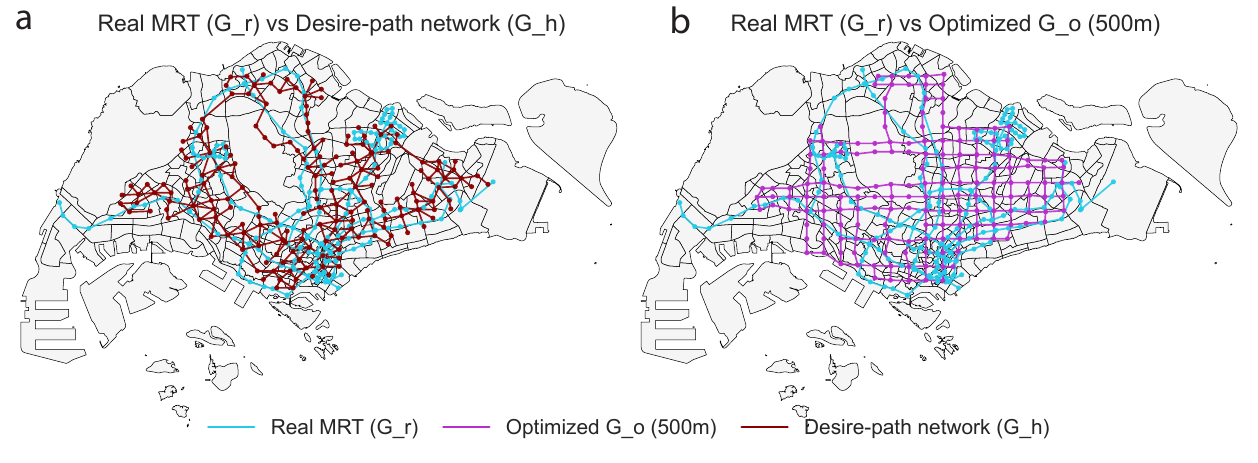}
    \caption{Comparison of benchmark and MRT networks in Singapore. Left: desire-path network (red) overlaid with the MRT network (cyan); mean Hausdorff distance: 5.49 km, and OT cost: 1.39 km. Right: system-optimal network (purple) overlaid with the MRT network (black); mean Hausdorff distance: 6.97 km, and OT cost: 2.12 km.}
    \label{fig:comp}
\end{figure}

As shown in Fig.~\ref{fig:comp}, the three networks exhibit broadly similar geometric silhouettes. Pronounced deviations between the benchmark and real-world networks are observed in central Singapore, around the city's largest nature reserve (Central Catchment Nature Reserve), where rail alignment is constrained by environmental protection. Notably, the system-optimal benchmark suggests several direct connections across this region but lacks the northern bypass that exists in both the real MRT and the desire-path network. The desire-path benchmark, by contrast, naturally avoids the reserve and more closely reproduces the northern routing of the real-world network. This reflects how each benchmark uses the OD data: the desire-path benchmark forms corridors only where OD trajectories concentrate, while the system-optimal benchmark places links anywhere on the grid that reduces aggregate travel time, regardless of trajectory density. While rail expansion across the nature reserve is challenging due to environmental protection, the system-optimal benchmark points to a potential network efficiency gain associated with improved central connectivity. We return to the practical interpretation of such signals in Section~6, where we discuss the implications of system-optimal versus desire-path planning extensions.

Quantitatively, the mean Hausdorff distance between the MRT and the desire-path benchmark is 5.49 km, while the corresponding distance between the MRT and the system-optimal benchmark is 6.97 km. Beyond geometric similarity, we compare the networks from a functional distribution perspective using optimal transport (OT). Based on the MRT OD flows, we compute OT costs that quantify the minimum spatial reallocation of interaction mass required to transform the real-world MRT network into either the desire-path or the system-optimal benchmark network \citep{peyre2019computational}. The corresponding OT costs (optimal objective values) are 1.39\,km (MRT versus desire-path network) and 2.12\,km (MRT versus system-optimal network). Both geometric similarity and OT measures imply that Singapore's MRT network is closer to the desire-path benchmark than to the system-optimal benchmark. In other words, the structural modifications required to steer the current system toward the desire-path configuration are smaller than those required to achieve the system-optimal configuration.

To assess whether the geographic constraints discussed above affect our conclusions, we additionally constructed a system-optimal network in which the Central Catchment Nature Reserve is treated as an unbuildable region. The qualitative result is unchanged: the empirical MRT network remains structurally closer to the desire-path benchmark than to this masked system-optimal benchmark (details in Appendix~C).

Furthermore, the desire-path and empirical MRT networks carry flow-based edge weights (passenger volumes), whereas the system-optimal network carries capacity-based weights (allocated travel speed). Under the system-optimal formulation, capacity is allocated in response to OD demand: links serving high-demand corridors receive higher speed or capacity. To verify that the OT comparison remains meaningful across these representations, we computed passenger flows on the system-optimal network by assigning OD trips to shortest-time paths under the optimized edge speeds. The resulting flow-based node strengths are strongly correlated with the capacity-based node strengths (details in Appendix~D), confirming that the spatial patterns captured by the OT comparison are consistent across both representations.

\section{Discussion and Conclusions}

In this work, we introduced a data-driven modeling framework to quantify similarities and discrepancies among three transportation network structures: a self-organized (or desire-path) network derived from local rules, a system-optimal network derived from global optimization, and the empirical real-world network. The desire-path benchmark captures two generic tendencies in navigation: a preference for near-direct travel and the emergence of shared high-use corridors driven by route consolidation and economies of scale. Applying the framework to Singapore, we find that its Mass Rapid Transit (MRT) network is substantially closer to the desire-path benchmark than to the system-optimal benchmark.

Our framework is intended as a descriptive benchmarking tool rather than a causal model of network formation. Both benchmark networks are derived (\enquote{reverse-engineered}) from observed OD patterns, which are themselves shaped by the existing MRT system. Because both benchmarks start from the same OD patterns, any influence of the existing network on these patterns affects them equally. The fact that one benchmark ends up much closer to the real MRT network than the other therefore reflects the two design logics themselves, not the shared OD input.

A tangible way to interpret the finding is to consider, as a conceptual thought experiment, what would happen if we rebuilt the network from scratch, given the current OD constellation. The desire-path benchmark represents a network that, if built, would align more closely with existing travel patterns. The system-optimal benchmark, by contrast, represents a structurally more different network that would more substantially alter accessibility across the city. Since changes in accessibility affect travel behavior, OD flows would be expected to more strongly adapt to the new network \citep{ortuzar2024modelling}. This points to a practical implication for infrastructure planning. When extending a transportation network based on a given OD constellation, as is often done in transportation planning practice, a desire-path approach may carry a lower risk of post-construction demand shifts, since the extension aligns with current travel patterns. In contrast, a system-optimal approach that targets efficiency gains for a given OD constellation can be expected to induce changes in the mobility flows away from what the design was optimized for. This resonates with recent arguments in urban science on the relative roles of top-down design and bottom-up adaptation in shaping urban systems \citep{bertaud2018order, bettencourt2021introduction, barthelemy2018morphogenesis}. In this view, successful urban structure emerges from decentralized adjustments to observed demand rather than from centrally optimized blueprints that may have unintended consequences. Nevertheless, integrating how OD patterns would shift following a metro-targeted intervention into our framework remains a direction for future work.

This finding also engages with a growing literature on metro network evolution. Longitudinal analyses of rail systems show that networks develop through phases of expansion, stagnation, and densification shaped by both spatial dynamics and planning interventions \citep{cats2017topological, pei2022efficiency}, with hierarchical progression from accessibility toward resilience and serviceability \citep{lin2022metro, yu2023urban}. Network expansion does not necessarily improve global efficiency, reflecting trade-offs between coverage, branching, and accessibility \citep{pei2022efficiency}. From a planning perspective, transport network design is typically framed as a multi-objective problem involving infrastructure, demand, and operational constraints \citep{wang2010global, jia2019review, shang2022integrated, duran2022survey}. Our results are consistent with this view: rather than approximating a single idealized model, the empirical MRT lies along a spectrum between demand-driven and efficiency-driven benchmarks, with a clear bias toward demand-driven structure.

Beyond these conceptual considerations, the two benchmarks provide complementary signals that can help prioritize practical planning interventions. Links that are consistently suggested by both the desire-path and system-optimal benchmarks, yet absent in the real-world network, may indicate high-impact connectivity gaps that jointly reflect latent demand and potential system-wide efficiency gains. Conversely, links that are not supported by either benchmark may indicate that they are weakly aligned with current OD demand. Such links may still be justified by objectives not captured in our framework (e.g., equity or long-term land-use development), and therefore may warrant closer interpretation. Links supported by only one benchmark require more careful evaluation: desire-path-only links may reflect strong localized demand and corridor formation, while system-optimal-only links may reflect network-wide efficiency improvements that do not emerge from local reinforcement alone. As discussed above, building such links may also shift OD flows, so their realized benefits may differ from those predicted under the current OD constellation. A concrete example is Singapore's Cross Island Line, which is currently under construction and includes tunnelling beneath the Central Catchment Nature Reserve \citep{LTA2022}: this corridor is flagged by the system-optimal benchmark but not by the desire-path benchmark, illustrating how system-optimal-only links may identify efficiency gains that justify infrastructure investment despite substantial local constraints. Importantly, these implications should be interpreted within the context of land-use, right-of-way feasibility, and environmental protection. In constrained areas (e.g., nature reserves), the choice between new construction and improvements along existing corridors depends on the magnitude of the efficiency gains and the feasibility of construction.

Our framework can be extended in several directions. First, practical constraints, such as protected areas, construction feasibility, and cost heterogeneity, could be incorporated directly into both the desire-path and system-optimal benchmarks, enabling a more realistic comparison to empirical systems. Second, the system-optimal formulation can be generalized to include additional objectives and constraints beyond aggregate travel time, such as energy consumption, accessibility, and robustness to disruptions. Third, OD patterns are influenced by the existing network, and future work could more explicitly account for the co-evolution of demand and infrastructure, for example, by iteratively updating OD flows under counterfactual network structures. Together, these extensions would further strengthen the utility of the proposed framework as a computational tool for diagnosing design-use misalignments and supporting adaptive transport planning.

\newpage

\appendix

\section*{Appendix}

\subsection*{Appendix A. Temporal stability of mobility patterns}
\label{appendix:mobility_recovery}
\renewcommand{\thefigure}{A\arabic{figure}}
\setcounter{figure}{0}

\begin{figure}[!b]
    \centering
    \includegraphics[width=0.95\linewidth, trim={0 0  0 0},clip]{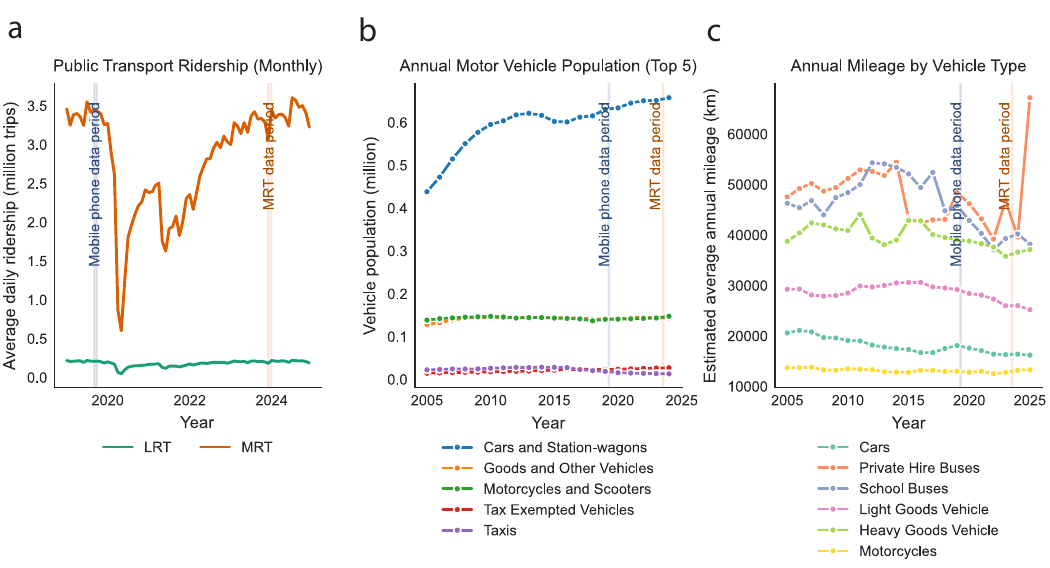}
    \caption{Mobility comparison between September 2019 and December 2023. Shaded bands indicate the two reference periods (i.e., \textit{mobile phone data period} and \textit{MRT data period}). a) Average daily ridership; b) Annual motor vehicle population by category; c) Annual private vehicle mileage by vehicle type. Data taken from Singapore's Land Transport Authority \citep{LTADataMall}.}
    \label{fig:mobility_comparison_2019_2023}
\end{figure}

To support the comparability of the datasets used in this study, we provide additional evidence for the stability of transport activity in Singapore. Figure~\ref{fig:mobility_comparison_2019_2023} summarizes key indicators derived from publicly available data sources. MRT ridership returned to pre-pandemic levels by 2023, while both vehicle population and mileage remained relatively stable over time. These patterns indicate that the overall scale and structure of urban mobility demand in Singapore remained broadly consistent between the two reference periods, i.e., September 2019 (mobile phone data) and December 2023 (MRT data).

\subsection*{Appendix B. Sensitivity analysis}
\label{appendix:Sensitivity analysis}
\renewcommand{\thefigure}{B\arabic{figure}}
\setcounter{figure}{0}
\renewcommand{\thetable}{B\arabic{table}}
\setcounter{table}{0}

To assess the robustness of our results, we conduct a comprehensive sensitivity analysis on the key model parameters, as summarized in Table~\ref{tab:parameters_sensitivity}. The table lists each parameter, its role in the model, the range of values tested, and the default setting used in the main analysis. Detailed results for each parameter are presented in the following subsections.

\begin{table}[!t]
    \centering
    \caption{Summary of key model parameters, their descriptions, and tested ranges (default values in \textbf{bold}).}
    \label{tab:parameters_sensitivity}
    \begin{tabular}{@{}>{\raggedright\arraybackslash}p{1.7cm}
                >{\raggedright\arraybackslash}p{7.5cm}
                >{\raggedright\arraybackslash}p{5.5cm}@{}}
        \toprule
        Parameter & Description & Tested range \\
        \midrule
        $h$ & KDE bandwidth (m) & 250, \textbf{500}, 1000, 2000, 3000 \\
        $T_d$ & Density threshold for node extraction (\%) & 5, \textbf{10}, 15, 20 \\
        $L$ & Grid cell size (m) & \textbf{500}, 750, 1000 \\
        \bottomrule
    \end{tabular}
\end{table}

\subsubsection*{KDE bandwidth ($h$)}
\label{appendix:bandwidth}

The KDE bandwidth $h$ controls the spatial range of attraction in the trajectory bundling process and therefore influences the formation of the network structure. In the main analysis, we set $h = 500$\,m to match the spatial resolution of the OD data. To assess robustness, we vary the bandwidth over a wide range, $h \in \{250, 500, 1000, 2000, 3000\}$\,m, and reconstruct the desire-path networks using the same procedure.

\begin{figure}[!t]
    \centering   
    \includegraphics[width=0.95\linewidth]{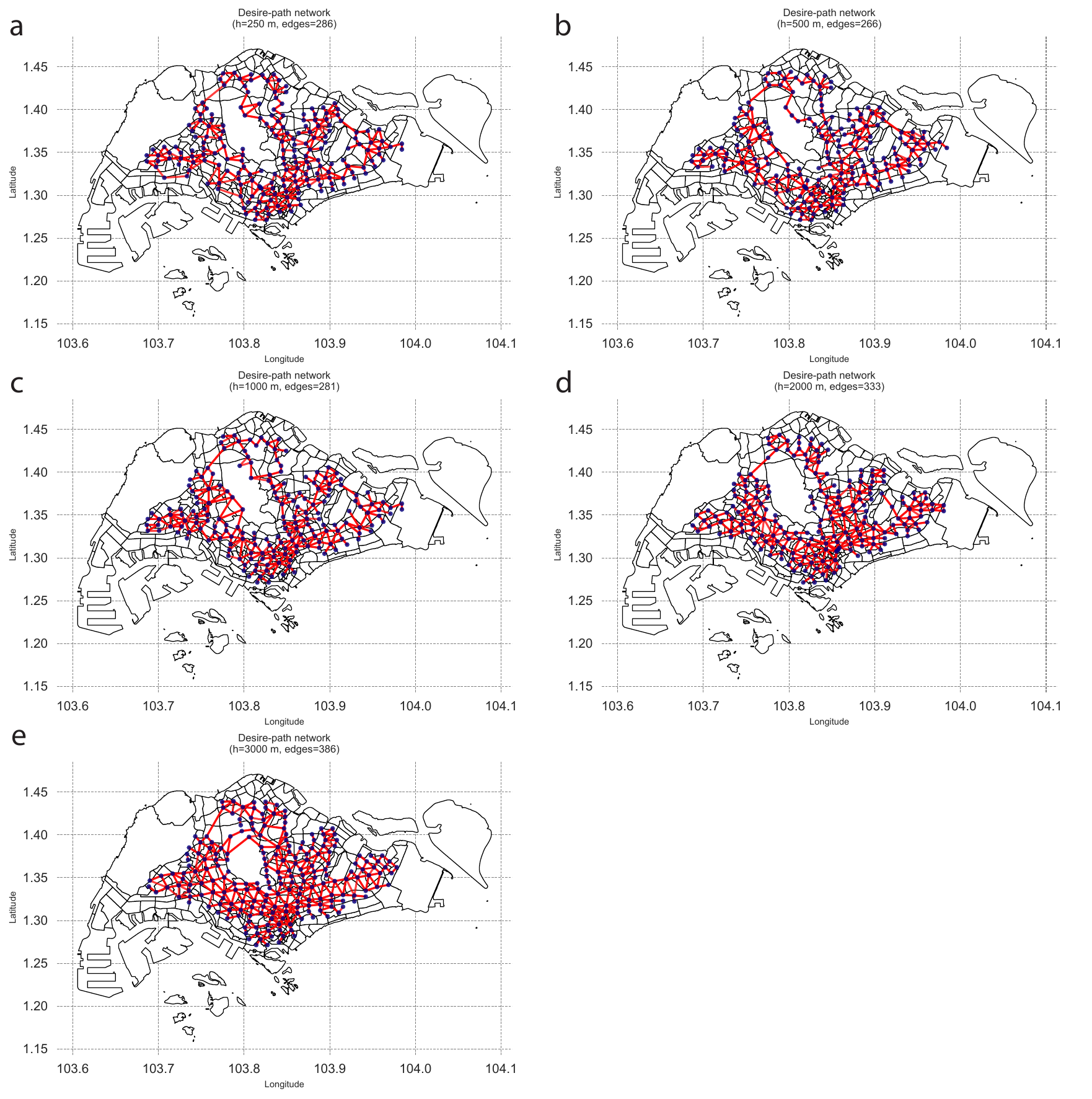}
    \caption{Desire-path network under different KDE bandwidths ($h$). a) $h = 250$ m; b) $h = 500$\,m; c) $h = 1000$\,m; d) $h = 2000$\,m; e) $h = 3000$\,m.}
    \label{fig: sensitive analysis h}
\end{figure}

\begin{figure}[!htbp]
    \centering
    \includegraphics[width=1\linewidth, trim={0 10 10 5}, clip]{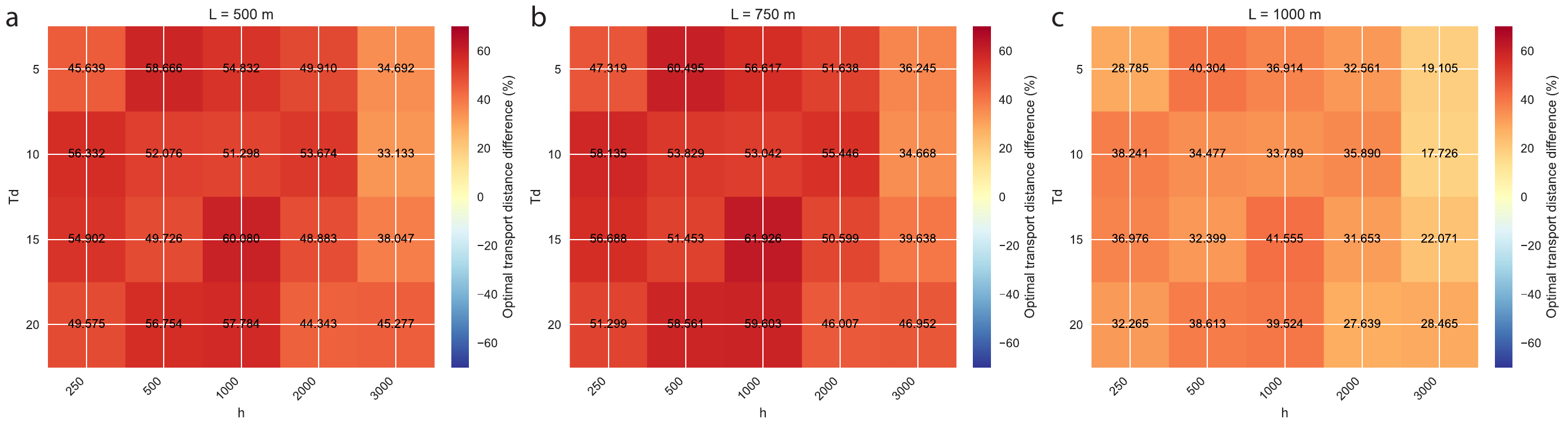}
    \caption{Comparison of the relative mismatch to the MRT network between the desire-path and system-optimal networks across tested parameter combinations. Each panel reports the percentage gap in optimal-transport distance, where positive values indicate that the desire-path network is closer to the MRT network than the system-optimal network. The percentage gap is computed as $(W_{G_r, G_o} - W_{G_r, G_h}) / W_{G_r, G_h} \times 100$, where $W_{G_r, G_o}$ and $W_{G_r, G_h}$ denote the OT distances from the real MRT network ($G_r$) to the system-optimal network ($G_o$) and to the desire-path network ($G_h$), respectively. a) $L = 500$\,m; b) $L = 750$\,m; c) $L = 1000$\,m.}
    \label{fig:sensitivity_comparison}
\end{figure}

The overall corridor patterns and extracted network topology remain qualitatively stable for bandwidths up to approximately 1\,km (Fig.~\ref{fig: sensitive analysis h}). Importantly, the relative comparison results remain consistent across all tested values, with the empirical MRT network being consistently closer to the desire-path benchmark than to the system-optimal benchmark (Fig.~\ref{fig:sensitivity_comparison}). These findings indicate that the main conclusions of this study are robust to the choice of KDE bandwidth within a reasonable range of spatial scales.

\subsubsection*{Density threshold ($T_d$)}
\label{appendix:density_threshold}

Desire-path network nodes are extracted from the KDE density surface to represent major mobility centers. The density threshold $T_d$ controls the level of aggregation: lower values retain more localized peaks, while higher values merge nearby centers and produce a more simplified network.

In practice, naive peak detection on the KDE surface produces many redundant local maxima. We therefore implement node identification as a three-stage procedure: an initial detection of local maxima above $T_d$ within a small neighborhood, an expanded scan with a larger neighborhood to capture additional regional peaks, and a pruning step that removes weak satellite peaks near stronger ones. The smaller neighborhood corresponds to the minimum inter-station spacing of the empirical MRT network, while the larger neighborhood corresponds to the maximum spacing, anchoring the procedure to the natural spatial scales of the MRT system.

\begin{figure}[!htbp]
    \centering
    \includegraphics[width=0.95\linewidth]{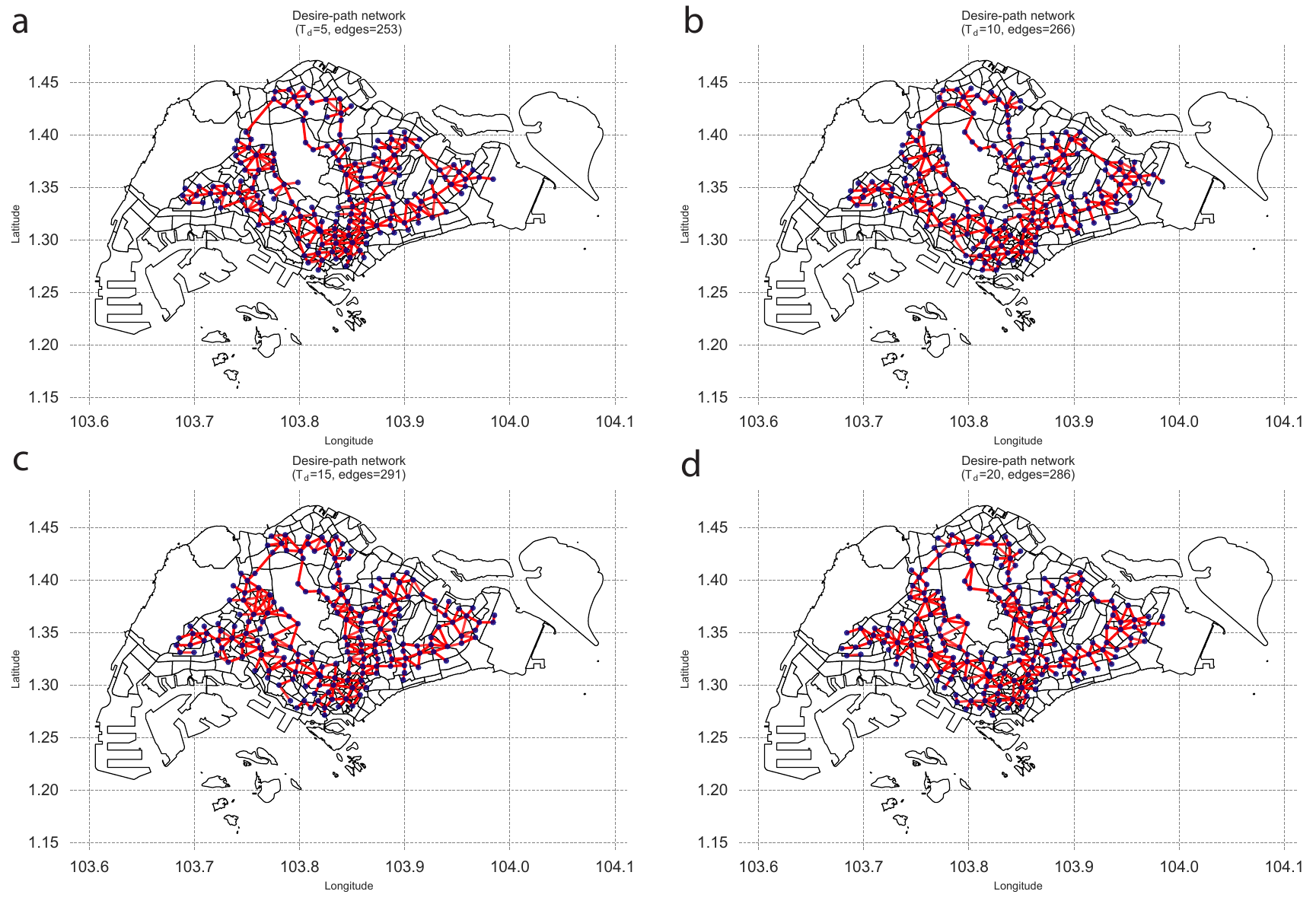}
    \caption{Desire-path network for different values of the density threshold for node extraction ($T_d$): a) $T_d = 5\%$; b) $T_d = 10\%$; c) $T_d = 15\%$; d) $T_d = 20\%$.}
    \label{fig:sensitivity_Td}
\end{figure}

We assess the sensitivity of the results by varying $T_d$ from 5\% to 20\%. While the number and spatial distribution of nodes change moderately (Fig.~\ref{fig:sensitivity_Td}), the overall network structure and the relative comparison results remain consistent. In particular, the empirical MRT network remains closer to the desire-path benchmark than to the system-optimal benchmark across all tested thresholds (Fig.~\ref{fig:sensitivity_comparison}). This confirms that the main conclusions are robust to the choice of density threshold.

\subsubsection*{Grid resolution in the system-optimal network ($L$)}
\label{appendix:sensitivity_grid}

To assess the sensitivity of the system-optimal network to spatial discretization, we conduct a sensitivity analysis by varying the grid resolution parameter $L$. In the main analysis, the system-optimal model is constructed on a spatial grid that defines the set of candidate nodes and links. The choice of grid resolution therefore determines the spatial granularity at which the network can be formed and may influence the resulting topology. We test three alternative resolutions, $L \in \{500, 750, 1000\}$\,m, covering a broad range of spatial scales relative to the empirical MRT network. For each resolution, the system-optimal network is reconstructed following the same optimization procedure.

\begin{figure}[!t]
    \centering
    \includegraphics[width=0.98\linewidth]{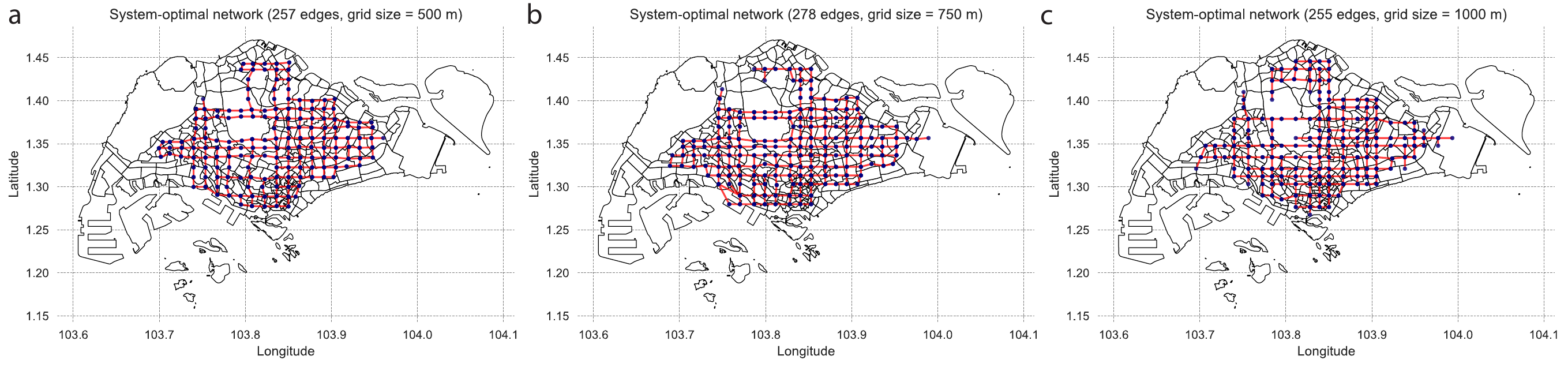}
    \caption{System-optimal network under different grid cell sizes ($L$): a) $L = 500$\,m; b) $L = 750$\,m; c) $L = 1000$\,m.}
    \label{fig:sensitivity_L}
\end{figure}

The resulting networks exhibit expected variations in spatial detail: finer grids allow more flexible link configurations and produce more locally adaptive structures, while coarser grids lead to more aggregated and simplified network layouts (Fig.~\ref{fig:sensitivity_L}). Despite these differences, the overall structural patterns remain consistent across all tested resolutions.

To quantify the impact on the network comparison, we recompute the optimal transport (OT) distances between the empirical MRT network and the benchmark networks for each grid resolution. The results (Fig.~\ref{fig:sensitivity_comparison}) show that the OT distance between the empirical MRT network and the desire-path benchmark ($W_{G_r, G_h}$) is consistently smaller than that to the system-optimal benchmark ($W_{G_r, G_o}$) across all tested resolutions. While spatial discretization affects local network configuration, it does not alter the key qualitative conclusions: the comparison between demand-driven and efficiency-driven network structures is robust to the choice of grid resolution.

\subsection*{Appendix C. Incorporating geographic constraints via spatial mask}
\label{appendix:spatial_mask}
\renewcommand{\thefigure}{C\arabic{figure}}
\setcounter{figure}{0}

The benchmark networks in the main analysis are intentionally constructed without explicit spatial constraints in order to isolate the underlying generative principles. However, real-world transport networks are subject to geographic and environmental restrictions. To assess their influence, we introduce a constrained scenario in which the Central Catchment Nature Reserve is treated as an unbuildable region.

\begin{figure}[!t]
    \centering
    \includegraphics[width=0.95\linewidth]{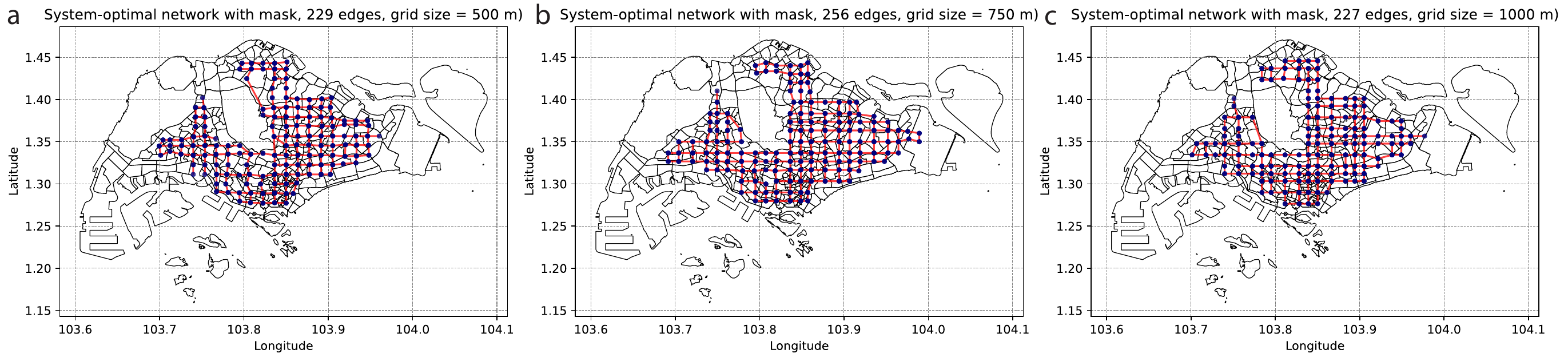}
        \caption{System-optimal network with a spatial mask preventing links from crossing the Central Catchment Nature Reserve, shown for different grid cell sizes ($L$). a) $L = 500$\,m; b) $L = 750$\,m; c) $L = 1000$\,m. Results across resolutions confirm that the masked network structure is largely robust to grid choice.}
    \label{fig: System-optimal network with spatial mask}
\end{figure}

For the system-optimal network, the spatial mask is implemented by removing grid cells and edges that overlap with the masked region, preventing direct connections across the Central Catchment Nature Reserve. The resulting system-optimal network (Fig.~\ref{fig: System-optimal network with spatial mask}) no longer contains direct links across the central region. Note that no direct connection is observed in the final desire-path network shown in Fig.~\ref{self}, so we omit the process here.

\begin{figure}[!t]
    \centering
    \includegraphics[width=1\linewidth, trim={0 10 10 0}, clip]{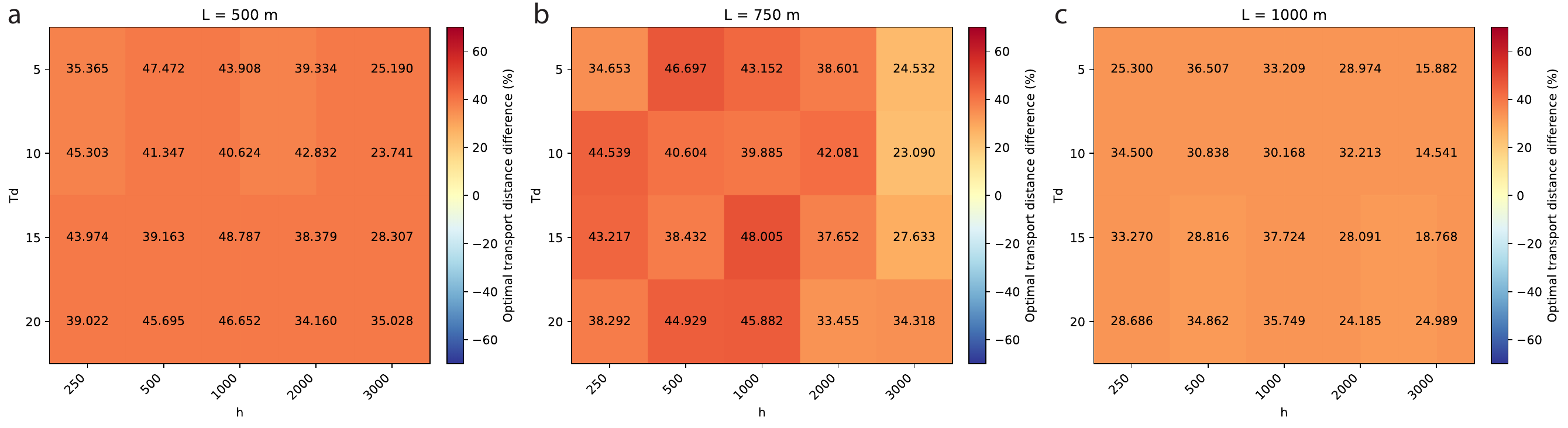}
    \caption{Comparison of the relative mismatch to the MRT network between the desire-path and system-optimal networks with spatial mask across tested parameter combinations. Each panel reports the percentage gap in optimal-transport distance, where positive values indicate that the desire-path network is closer to the MRT network than the system-optimal network. The percentage gap is computed as $(W_{G_r, G_o} - W_{G_r, G_h}) / W_{G_r, G_h} \times 100$, where $W_{G_r, G_o}$ and $W_{G_r, G_h}$ denote the OT distances from the real MRT network ($G_r$) to the system-optimal network ($G_o$) and to the desire-path network ($G_h$), respectively. a) $L = 500$\,m; b) $L = 750$\,m; c) $L = 1000$\,m.}
    \label{fig: sensitivity analysis comparison with spatial mask for optimal network}
\end{figure}

Recomputing the OT distances under this constrained scenario (Fig.~\ref{fig: sensitivity analysis comparison with spatial mask for optimal network}) shows that the absolute discrepancy between the empirical network and the system-optimal benchmark decreases (compare to Fig.~\ref{fig:sensitivity_comparison}), but the qualitative ranking is preserved: the empirical MRT network remains structurally closer to the desire-path benchmark than to the system-optimal benchmark. This suggests that, at least for the case of Singapore's Central Catchment Nature Reserve, geographic constraints primarily affect local network configuration without altering the overall structural ranking of the benchmarks.

\subsection*{Appendix D. Supply-demand alignment in the system-optimal network}
\renewcommand{\thefigure}{D\arabic{figure}}
\setcounter{figure}{0}
The OT comparison uses node strength as a dimensionless proxy for location importance. In the desire-path and empirical MRT networks, node strength is computed from passenger flows, while in the system-optimal network it is computed from allocated capacities. To verify that these two representations yield comparable spatial distributions, we computed passenger flows on the system-optimal network by assigning OD trips to shortest-time paths under the optimized edge speeds. Figure~\ref{fig:supply_demand_correlation} shows the resulting flow distribution against the capacity distribution. The two distributions are strongly correlated (Pearson's $r$=0.975, Spearman's $\rho$=0.991), confirming that the spatial patterns captured by the OT comparison are consistent across demand-based and capacity-based representations.
\begin{figure}[!htbp]
    \centering
    \includegraphics[width=0.50\linewidth]{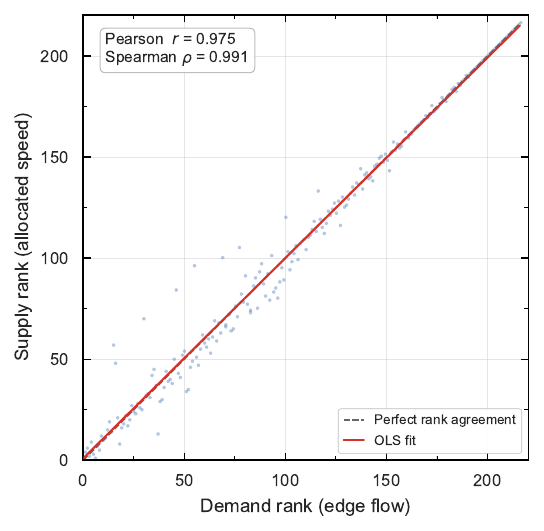}
    \caption{Correlation between supply (allocated capacity) and demand (passenger flows) in the system-optimal network. Each point corresponds to an edge, ranked by its flow/capacity value. Pearson's $r$ = 0.975, Spearman's $\rho$ = 0.991.}
    \label{fig:supply_demand_correlation}   
\end{figure}

\newpage
\section*{Author contributions}

\textbf{Tianyu Dong:} Conceptualization, Methodology, Formal analysis, Data Curation, Writing - Original Draft. \textbf{Jiazu Zhou:} Conceptualization, Methodology, Formal analysis, Data Curation, Writing - Review $\&$ Editing. \textbf{Markus Schläpfer:} Conceptualization, Writing - Review $\&$ Editing, Supervision.

\section*{Acknowledgments}

M.S. acknowledges support from the start-up funds provided by Columbia University. Part of this research was conducted at the Singapore-ETH Centre, which is supported and funded by the National Research Foundation and ETH Zurich, with contributions from the National University of Singapore, Nanyang Technological University and the Singapore University of Technology and Design.

\section*{Declaration of generative AI and AI-assisted technologies in the manuscript preparation process}

During the preparation of this work the authors used ChatGPT (version 5.2) from OpenAI in order to improve language and readability. After using this tool, the authors reviewed and edited the content as needed and take full responsibility for the content of the published article.

\bibliographystyle{elsarticle-harv}
\bibliography{references}

\begin{thebibliography}{62}
\expandafter\ifx\csname natexlab\endcsname\relax\def\natexlab#1{#1}\fi
\providecommand{\url}[1]{\texttt{#1}}
\providecommand{\href}[2]{#2}
\providecommand{\path}[1]{#1}
\providecommand{\DOIprefix}{doi:}
\providecommand{\ArXivprefix}{arXiv:}
\providecommand{\URLprefix}{URL: }
\providecommand{\Pubmedprefix}{pmid:}
\providecommand{\doi}[1]{\href{http://dx.doi.org/#1}{\path{#1}}}
\providecommand{\Pubmed}[1]{\href{pmid:#1}{\path{#1}}}
\providecommand{\bibinfo}[2]{#2}
\ifx\xfnm\relax \def\xfnm[#1]{\unskip,\space#1}\fi
%Type = Article
\bibitem[{Ahern et~al.(2022)Ahern, Paz and Corry}]{ahern2022approximate}
\bibinfo{author}{Ahern, Z.}, \bibinfo{author}{Paz, A.}, \bibinfo{author}{Corry,
  P.}, \bibinfo{year}{2022}.
\newblock \bibinfo{title}{Approximate multi-objective optimization for
  integrated bus route design and service frequency setting}.
\newblock \bibinfo{journal}{Transportation Research Part B: Methodological}
  \bibinfo{volume}{155}, \bibinfo{pages}{1--25}.
%Type = Article
\bibitem[{Alessandretti et~al.(2023)Alessandretti, Orozco, Saberi, Szell and
  Battiston}]{alessandretti2023}
\bibinfo{author}{Alessandretti, L.}, \bibinfo{author}{Orozco, L.G.N.},
  \bibinfo{author}{Saberi, M.}, \bibinfo{author}{Szell, M.},
  \bibinfo{author}{Battiston, F.}, \bibinfo{year}{2023}.
\newblock \bibinfo{title}{Multimodal urban mobility and multilayer transport
  networks}.
\newblock \bibinfo{journal}{Environment and Planning B: Urban Analytics and
  City Science} \bibinfo{volume}{50}, \bibinfo{pages}{2038--2070}.
%Type = Book
\bibitem[{Barth{\'e}lemy(2018)}]{barthelemy2018morphogenesis}
\bibinfo{author}{Barth{\'e}lemy, M.}, \bibinfo{year}{2018}.
\newblock \bibinfo{title}{Morphogenesis of spatial networks}.
\newblock \bibinfo{publisher}{Springer}.
%Type = Article
\bibitem[{Batty(2008)}]{batty2008size}
\bibinfo{author}{Batty, M.}, \bibinfo{year}{2008}.
\newblock \bibinfo{title}{The size, scale, and shape of cities}.
\newblock \bibinfo{journal}{Science} \bibinfo{volume}{319},
  \bibinfo{pages}{769--771}.
%Type = Book
\bibitem[{Batty(2013)}]{batty2013new}
\bibinfo{author}{Batty, M.}, \bibinfo{year}{2013}.
\newblock \bibinfo{title}{The new science of cities}.
\newblock \bibinfo{publisher}{MIT Press}.
%Type = Incollection
\bibitem[{Batty(2016)}]{batty2016complexity}
\bibinfo{author}{Batty, M.}, \bibinfo{year}{2016}.
\newblock \bibinfo{title}{Complexity in city systems: Understanding, evolution,
  and design}, in: \bibinfo{booktitle}{A planner's encounter with complexity}.
  \bibinfo{publisher}{Routledge}, pp. \bibinfo{pages}{99--122}.
%Type = Article
\bibitem[{Beaudoin et~al.(2015)Beaudoin, Farzin and
  Lawell}]{beaudoin2015public}
\bibinfo{author}{Beaudoin, J.}, \bibinfo{author}{Farzin, Y.H.},
  \bibinfo{author}{Lawell, C.Y.C.L.}, \bibinfo{year}{2015}.
\newblock \bibinfo{title}{Public transit investment and sustainable
  transportation: A review of studies of transit's impact on traffic congestion
  and air quality}.
\newblock \bibinfo{journal}{Research in Transportation Economics}
  \bibinfo{volume}{52}, \bibinfo{pages}{15--22}.
%Type = Book
\bibitem[{Ben-Akiva and Lerman(1985)}]{ben1985discrete}
\bibinfo{author}{Ben-Akiva, M.E.}, \bibinfo{author}{Lerman, S.R.},
  \bibinfo{year}{1985}.
\newblock \bibinfo{title}{Discrete choice analysis: theory and application to
  travel demand}.
\newblock \bibinfo{publisher}{MIT Press}.
%Type = Book
\bibitem[{Bertaud(2018)}]{bertaud2018order}
\bibinfo{author}{Bertaud, A.}, \bibinfo{year}{2018}.
\newblock \bibinfo{title}{Order without design: How markets shape cities}.
\newblock \bibinfo{publisher}{MIT Press}.
%Type = Book
\bibitem[{Bettencourt(2021)}]{bettencourt2021introduction}
\bibinfo{author}{Bettencourt, L.M.}, \bibinfo{year}{2021}.
\newblock \bibinfo{title}{Introduction to urban science: {E}vidence and theory
  of cities as complex systems}.
\newblock \bibinfo{publisher}{MIT Press}.
%Type = Article
\bibitem[{Bontorin et~al.(2024)Bontorin, Cencetti, Gallotti, Lepri and
  De~Domenico}]{bontorin2024emergence}
\bibinfo{author}{Bontorin, S.}, \bibinfo{author}{Cencetti, G.},
  \bibinfo{author}{Gallotti, R.}, \bibinfo{author}{Lepri, B.},
  \bibinfo{author}{De~Domenico, M.}, \bibinfo{year}{2024}.
\newblock \bibinfo{title}{Emergence of complex network topologies from
  flow-weighted optimization of network efficiency}.
\newblock \bibinfo{journal}{Physical Review X} \bibinfo{volume}{14},
  \bibinfo{pages}{021050}.
%Type = Article
\bibitem[{Cats(2017)}]{cats2017topological}
\bibinfo{author}{Cats, O.}, \bibinfo{year}{2017}.
\newblock \bibinfo{title}{Topological evolution of a metropolitan rail
  transport network: The case of {S}tockholm}.
\newblock \bibinfo{journal}{Journal of Transport Geography}
  \bibinfo{volume}{62}, \bibinfo{pages}{172--183}.
%Type = Misc
\bibitem[{{CityData.AI}(2025)}]{CityDataAI}
\bibinfo{author}{{CityData.AI}}, \bibinfo{year}{2025}.
\newblock \bibinfo{note}{Available at: \url{https://citydata.ai}}.
%Type = Article
\bibitem[{Diao et~al.(2021)Diao, Kong and Zhao}]{diao2021impacts}
\bibinfo{author}{Diao, M.}, \bibinfo{author}{Kong, H.}, \bibinfo{author}{Zhao,
  J.}, \bibinfo{year}{2021}.
\newblock \bibinfo{title}{Impacts of transportation network companies on urban
  mobility}.
\newblock \bibinfo{journal}{Nature Sustainability} \bibinfo{volume}{4},
  \bibinfo{pages}{494--500}.
%Type = Article
\bibitem[{Dur{\'a}n-Micco and Vansteenwegen(2022)}]{duran2022survey}
\bibinfo{author}{Dur{\'a}n-Micco, J.}, \bibinfo{author}{Vansteenwegen, P.},
  \bibinfo{year}{2022}.
\newblock \bibinfo{title}{A survey on the transit network design and frequency
  setting problem}.
\newblock \bibinfo{journal}{Public transport} \bibinfo{volume}{14},
  \bibinfo{pages}{155--190}.
%Type = Book
\bibitem[{Erlander and Stewart(1990)}]{erlander1990gravity}
\bibinfo{author}{Erlander, S.}, \bibinfo{author}{Stewart, N.F.},
  \bibinfo{year}{1990}.
\newblock \bibinfo{title}{The gravity model in transportation analysis: Theory
  and extensions}.
\newblock \bibinfo{publisher}{VSP}.
%Type = Article
\bibitem[{Farahani et~al.(2013)Farahani, Miandoabchi, Szeto and
  Rashidi}]{farahani2013review}
\bibinfo{author}{Farahani, R.Z.}, \bibinfo{author}{Miandoabchi, E.},
  \bibinfo{author}{Szeto, W.Y.}, \bibinfo{author}{Rashidi, H.},
  \bibinfo{year}{2013}.
\newblock \bibinfo{title}{A review of urban transportation network design
  problems}.
\newblock \bibinfo{journal}{European Journal of Operational Research}
  \bibinfo{volume}{229}, \bibinfo{pages}{281--302}.
%Type = Article
\bibitem[{Farber and Fu(2017)}]{FARBER201730}
\bibinfo{author}{Farber, S.}, \bibinfo{author}{Fu, L.}, \bibinfo{year}{2017}.
\newblock \bibinfo{title}{Dynamic public transit accessibility using travel
  time cubes: Comparing the effects of infrastructure (dis)investments over
  time}.
\newblock \bibinfo{journal}{Computers, Environment and Urban Systems}
  \bibinfo{volume}{62}, \bibinfo{pages}{30--40}.
%Type = Article
\bibitem[{Foster and Newell(2019)}]{foster2019detroit}
\bibinfo{author}{Foster, A.}, \bibinfo{author}{Newell, J.P.},
  \bibinfo{year}{2019}.
\newblock \bibinfo{title}{Detroit's lines of desire: Footpaths and vacant land
  in the motor city}.
\newblock \bibinfo{journal}{Landscape and Urban Planning}
  \bibinfo{volume}{189}, \bibinfo{pages}{260--273}.
%Type = Article
\bibitem[{Gerrits et~al.(2024)Gerrits, van Heeswijk and
  Mes}]{gerrits2024towards}
\bibinfo{author}{Gerrits, B.}, \bibinfo{author}{van Heeswijk, W.},
  \bibinfo{author}{Mes, M.}, \bibinfo{year}{2024}.
\newblock \bibinfo{title}{Towards self-organizing logistics in transportation:
  A literature review and typology}.
\newblock \bibinfo{journal}{International Transactions in Operational Research}
  \bibinfo{volume}{31}, \bibinfo{pages}{1309--1374}.
%Type = Article
\bibitem[{Han et~al.(2015)Han, Friesz, Szeto and Liu}]{han2015elastic}
\bibinfo{author}{Han, K.}, \bibinfo{author}{Friesz, T.L.},
  \bibinfo{author}{Szeto, W.}, \bibinfo{author}{Liu, H.}, \bibinfo{year}{2015}.
\newblock \bibinfo{title}{Elastic demand dynamic network user equilibrium:
  Formulation, existence and computation}.
\newblock \bibinfo{journal}{Transportation Research Part B: Methodological}
  \bibinfo{volume}{81}, \bibinfo{pages}{183--209}.
%Type = Article
\bibitem[{Hasnine and Habib(2018)}]{hasnine2018dynamics}
\bibinfo{author}{Hasnine, M.S.}, \bibinfo{author}{Habib, K.N.},
  \bibinfo{year}{2018}.
\newblock \bibinfo{title}{What about the dynamics in daily travel mode choices?
  {A} dynamic discrete choice approach for tour-based mode choice modelling}.
\newblock \bibinfo{journal}{Transport Policy} \bibinfo{volume}{71},
  \bibinfo{pages}{70--80}.
%Type = Article
\bibitem[{Helbing et~al.(1997)Helbing, Keltsch and
  Molnar}]{helbing1997modelling}
\bibinfo{author}{Helbing, D.}, \bibinfo{author}{Keltsch, J.},
  \bibinfo{author}{Molnar, P.}, \bibinfo{year}{1997}.
\newblock \bibinfo{title}{Modelling the evolution of human trail systems}.
\newblock \bibinfo{journal}{Nature} \bibinfo{volume}{388},
  \bibinfo{pages}{47--50}.
%Type = Inproceedings
\bibitem[{Holten and Van~Wijk(2009)}]{holten2009force}
\bibinfo{author}{Holten, D.}, \bibinfo{author}{Van~Wijk, J.J.},
  \bibinfo{year}{2009}.
\newblock \bibinfo{title}{Force-directed edge bundling for graph
  visualization}, in: \bibinfo{booktitle}{Computer graphics forum},
  \bibinfo{organization}{Wiley Online Library}. pp. \bibinfo{pages}{983--990}.
%Type = Article
\bibitem[{Hosseininasab et~al.(2018)Hosseininasab, Shetab-Boushehri, Hejazi and
  Karimi}]{hosseininasab2018multi}
\bibinfo{author}{Hosseininasab, S.M.}, \bibinfo{author}{Shetab-Boushehri,
  S.N.}, \bibinfo{author}{Hejazi, S.R.}, \bibinfo{author}{Karimi, H.},
  \bibinfo{year}{2018}.
\newblock \bibinfo{title}{A multi-objective integrated model for selecting,
  scheduling, and budgeting road construction projects}.
\newblock \bibinfo{journal}{European Journal of Operational Research}
  \bibinfo{volume}{271}, \bibinfo{pages}{262--277}.
%Type = Article
\bibitem[{Huttenlocher et~al.(1993)Huttenlocher, Klanderman and
  Rucklidge}]{huttenlocher1993comparing}
\bibinfo{author}{Huttenlocher, D.P.}, \bibinfo{author}{Klanderman, G.A.},
  \bibinfo{author}{Rucklidge, W.J.}, \bibinfo{year}{1993}.
\newblock \bibinfo{title}{Comparing images using the {H}ausdorff distance}.
\newblock \bibinfo{journal}{IEEE Transactions on Pattern Analysis and Machine
  Intelligence} \bibinfo{volume}{15}, \bibinfo{pages}{850--863}.
%Type = Article
\bibitem[{Jia et~al.(2019)Jia, Ma and Hu}]{jia2019review}
\bibinfo{author}{Jia, G.L.}, \bibinfo{author}{Ma, R.G.}, \bibinfo{author}{Hu,
  Z.H.}, \bibinfo{year}{2019}.
\newblock \bibinfo{title}{Review of urban transportation network design
  problems based on citespace}.
\newblock \bibinfo{journal}{Mathematical Problems in Engineering}
  \bibinfo{volume}{2019}, \bibinfo{pages}{5735702}.
%Type = Article
\bibitem[{Jiang et~al.(2017)Jiang, Ferreira and Gonzalez}]{jiang2017activity}
\bibinfo{author}{Jiang, S.}, \bibinfo{author}{Ferreira, J.},
  \bibinfo{author}{Gonzalez, M.C.}, \bibinfo{year}{2017}.
\newblock \bibinfo{title}{Activity-based human mobility patterns inferred from
  mobile phone data: A case study of {S}ingapore}.
\newblock \bibinfo{journal}{IEEE Transactions on Big Data} \bibinfo{volume}{3},
  \bibinfo{pages}{208--219}.
%Type = Article
\bibitem[{Keller and Gavrila(2013)}]{keller2013will}
\bibinfo{author}{Keller, C.G.}, \bibinfo{author}{Gavrila, D.M.},
  \bibinfo{year}{2013}.
\newblock \bibinfo{title}{Will the pedestrian cross? {A} study on pedestrian
  path prediction}.
\newblock \bibinfo{journal}{IEEE Transactions on Intelligent Transportation
  Systems} \bibinfo{volume}{15}, \bibinfo{pages}{494--506}.
%Type = Article
\bibitem[{Kujala et~al.(2018)Kujala, Weckstr\"om, Mladenovi\'c and
  Saram\"aki}]{kujala2018travel}
\bibinfo{author}{Kujala, R.}, \bibinfo{author}{Weckstr\"om, C.},
  \bibinfo{author}{Mladenovi\'c, M.N.}, \bibinfo{author}{Saram\"aki, J.},
  \bibinfo{year}{2018}.
\newblock \bibinfo{title}{Travel times and transfers in public transport:
  Comprehensive accessibility analysis based on pareto-optimal journeys}.
\newblock \bibinfo{journal}{Computers, Environment and Urban Systems}
  \bibinfo{volume}{67}, \bibinfo{pages}{41--54}.
%Type = Misc
\bibitem[{{Land Transport Authority Singapore}(2022)}]{LTA2022}
\bibinfo{author}{{Land Transport Authority Singapore}}, \bibinfo{year}{2022}.
\newblock \bibinfo{title}{Lta annual report 2021/22}.
\newblock \bibinfo{note}{Available at:
  \url{https://www.lta.gov.sg/content/dam/ltagov/who_we_are/statistics_and_publications/report/pdf/LTA_AR2122.pdf}}.
%Type = Misc
\bibitem[{{Land Transport Authority Singapore}(2024)}]{LTADataMall}
\bibinfo{author}{{Land Transport Authority Singapore}}, \bibinfo{year}{2024}.
\newblock \bibinfo{title}{{LTA} {DataMall}}.
\newblock \bibinfo{note}{Available at:
  \url{https://datamall.lta.gov.sg/content/datamall/en.html}}.
%Type = Inproceedings
\bibitem[{Lee et~al.(2007)Lee, Han and Whang}]{lee2007trajectory}
\bibinfo{author}{Lee, J.G.}, \bibinfo{author}{Han, J.}, \bibinfo{author}{Whang,
  K.Y.}, \bibinfo{year}{2007}.
\newblock \bibinfo{title}{Trajectory clustering: a partition-and-group
  framework}, in: \bibinfo{booktitle}{Proceedings of the 2007 ACM SIGMOD
  international conference on Management of data}, pp.
  \bibinfo{pages}{593--604}.
%Type = Article
\bibitem[{Leite and De~Bacco(2024)}]{leite2024simopt}
\bibinfo{author}{Leite, D.}, \bibinfo{author}{De~Bacco, C.},
  \bibinfo{year}{2024}.
\newblock \bibinfo{title}{Similarity and economy of scale in urban
  transportation networks and optimal transport-based infrastructures}.
\newblock \bibinfo{journal}{Nature Communications} \bibinfo{volume}{15},
  \bibinfo{pages}{7981}.
%Type = Article
\bibitem[{Levinson and Yerra(2006)}]{levinson2006self}
\bibinfo{author}{Levinson, D.}, \bibinfo{author}{Yerra, B.},
  \bibinfo{year}{2006}.
\newblock \bibinfo{title}{Self-organization of surface transportation
  networks}.
\newblock \bibinfo{journal}{Transportation Science} \bibinfo{volume}{40},
  \bibinfo{pages}{179--188}.
%Type = Article
\bibitem[{L{\'e}vy and Schwindt(2018)}]{levy2018notions}
\bibinfo{author}{L{\'e}vy, B.}, \bibinfo{author}{Schwindt, E.L.},
  \bibinfo{year}{2018}.
\newblock \bibinfo{title}{Notions of optimal transport theory and how to
  implement them on a computer}.
\newblock \bibinfo{journal}{Computers \& Graphics} \bibinfo{volume}{72},
  \bibinfo{pages}{135--148}.
%Type = Book
\bibitem[{Lidwell et~al.(2010)Lidwell, Holden and
  Butler}]{lidwell2010universal}
\bibinfo{author}{Lidwell, W.}, \bibinfo{author}{Holden, K.},
  \bibinfo{author}{Butler, J.}, \bibinfo{year}{2010}.
\newblock \bibinfo{title}{Universal principles of design, revised and updated:
  125 ways to enhance usability, influence perception, increase appeal, make
  better design decisions, and teach through design}.
\newblock \bibinfo{publisher}{Rockport Publishers}.
%Type = Article
\bibitem[{Lin et~al.(2022)Lin, Broere and Cui}]{lin2022metro}
\bibinfo{author}{Lin, D.}, \bibinfo{author}{Broere, W.}, \bibinfo{author}{Cui,
  J.}, \bibinfo{year}{2022}.
\newblock \bibinfo{title}{Metro systems and urban development: Impacts and
  implications}.
\newblock \bibinfo{journal}{Tunnelling and underground space technology}
  \bibinfo{volume}{125}, \bibinfo{pages}{104509}.
%Type = Article
\bibitem[{Liu et~al.(2020)Liu, Feng, Zhang, Hua and Li}]{liu2020pareto}
\bibinfo{author}{Liu, Y.}, \bibinfo{author}{Feng, X.}, \bibinfo{author}{Zhang,
  L.}, \bibinfo{author}{Hua, W.}, \bibinfo{author}{Li, K.},
  \bibinfo{year}{2020}.
\newblock \bibinfo{title}{A {P}areto artificial fish swarm algorithm for
  solving a multi-objective electric transit network design problem}.
\newblock \bibinfo{journal}{Transportmetrica A: Transport Science}
  \bibinfo{volume}{16}, \bibinfo{pages}{1648--1670}.
%Type = Article
\bibitem[{Luathep et~al.(2011)Luathep, Sumalee, Lam, Li and
  Lo}]{luathep2011global}
\bibinfo{author}{Luathep, P.}, \bibinfo{author}{Sumalee, A.},
  \bibinfo{author}{Lam, W.H.}, \bibinfo{author}{Li, Z.C.}, \bibinfo{author}{Lo,
  H.K.}, \bibinfo{year}{2011}.
\newblock \bibinfo{title}{Global optimization method for mixed transportation
  network design problem: {A} mixed-integer linear programming approach}.
\newblock \bibinfo{journal}{Transportation Research Part B: Methodological}
  \bibinfo{volume}{45}, \bibinfo{pages}{808--827}.
%Type = Article
\bibitem[{{McFadden}(2007)}]{mcfaddenc2007behavioral}
\bibinfo{author}{{McFadden}, D.}, \bibinfo{year}{2007}.
\newblock \bibinfo{title}{The behavioral science of transportation}.
\newblock \bibinfo{journal}{Transport Policy} \bibinfo{volume}{14},
  \bibinfo{pages}{269--274}.
%Type = Article
\bibitem[{Meyur et~al.(2022)Meyur, Vullikanti, Swarup, Mortveit, Centeno,
  Phadke, Poor and Marathe}]{meyur2022ensembles}
\bibinfo{author}{Meyur, R.}, \bibinfo{author}{Vullikanti, A.},
  \bibinfo{author}{Swarup, S.}, \bibinfo{author}{Mortveit, H.S.},
  \bibinfo{author}{Centeno, V.}, \bibinfo{author}{Phadke, A.},
  \bibinfo{author}{Poor, H.V.}, \bibinfo{author}{Marathe, M.V.},
  \bibinfo{year}{2022}.
\newblock \bibinfo{title}{Ensembles of realistic power distribution networks}.
\newblock \bibinfo{journal}{Proceedings of the National Academy of Sciences}
  \bibinfo{volume}{119}, \bibinfo{pages}{e2205772119}.
%Type = Article
\bibitem[{Mittal et~al.(2024)Mittal, Timme and
  Schr{\"o}der}]{mittal2024efficient}
\bibinfo{author}{Mittal, K.M.}, \bibinfo{author}{Timme, M.},
  \bibinfo{author}{Schr{\"o}der, M.}, \bibinfo{year}{2024}.
\newblock \bibinfo{title}{Efficient self-organization of informal public
  transport networks}.
\newblock \bibinfo{journal}{Nature Communications} \bibinfo{volume}{15},
  \bibinfo{pages}{4910}.
%Type = Article
\bibitem[{Mwale et~al.(2022)Mwale, Luke and Pisa}]{mwale2022factors}
\bibinfo{author}{Mwale, M.}, \bibinfo{author}{Luke, R.}, \bibinfo{author}{Pisa,
  N.}, \bibinfo{year}{2022}.
\newblock \bibinfo{title}{Factors that affect travel behaviour in developing
  cities: A methodological review}.
\newblock \bibinfo{journal}{Transportation Research Interdisciplinary
  Perspectives} \bibinfo{volume}{16}, \bibinfo{pages}{100683}.
%Type = Book
\bibitem[{Ort{\'u}zar and Willumsen(2024)}]{ortuzar2024modelling}
\bibinfo{author}{Ort{\'u}zar, J.d.D.}, \bibinfo{author}{Willumsen, L.G.},
  \bibinfo{year}{2024}.
\newblock \bibinfo{title}{Modelling transport}.
\newblock \bibinfo{publisher}{John Wiley \& Sons}.
%Type = Article
\bibitem[{Pappalardo et~al.(2023)Pappalardo, Manley, Sekara and
  Alessandretti}]{pappalardo2023future}
\bibinfo{author}{Pappalardo, L.}, \bibinfo{author}{Manley, E.},
  \bibinfo{author}{Sekara, V.}, \bibinfo{author}{Alessandretti, L.},
  \bibinfo{year}{2023}.
\newblock \bibinfo{title}{Future directions in human mobility science}.
\newblock \bibinfo{journal}{Nature Computational Science} \bibinfo{volume}{3},
  \bibinfo{pages}{588--600}.
%Type = Article
\bibitem[{Pei et~al.(2022)Pei, Xiao, Yu and Li}]{pei2022efficiency}
\bibinfo{author}{Pei, A.}, \bibinfo{author}{Xiao, F.}, \bibinfo{author}{Yu,
  S.}, \bibinfo{author}{Li, L.}, \bibinfo{year}{2022}.
\newblock \bibinfo{title}{Efficiency in the evolution of metro networks}.
\newblock \bibinfo{journal}{Scientific Reports} \bibinfo{volume}{12},
  \bibinfo{pages}{8326}.
%Type = Article
\bibitem[{Peyr{\'e} and Cuturi(2019)}]{peyre2019computational}
\bibinfo{author}{Peyr{\'e}, G.}, \bibinfo{author}{Cuturi, M.},
  \bibinfo{year}{2019}.
\newblock \bibinfo{title}{Computational optimal transport: {W}ith applications
  to data science}.
\newblock \bibinfo{journal}{Foundations and Trends in Machine Learning}
  \bibinfo{volume}{11}, \bibinfo{pages}{355--607}.
%Type = Article
\bibitem[{Pinto et~al.(2020)Pinto, Hyland, Mahmassani and
  Verbas}]{pinto2020joint}
\bibinfo{author}{Pinto, H.K.}, \bibinfo{author}{Hyland, M.F.},
  \bibinfo{author}{Mahmassani, H.S.}, \bibinfo{author}{Verbas, I.{\"O}.},
  \bibinfo{year}{2020}.
\newblock \bibinfo{title}{Joint design of multimodal transit networks and
  shared autonomous mobility fleets}.
\newblock \bibinfo{journal}{Transportation Research Part C: Emerging
  Technologies} \bibinfo{volume}{113}, \bibinfo{pages}{2--20}.
%Type = Book
\bibitem[{Santambrogio(2015)}]{santambrogio2015optimal}
\bibinfo{author}{Santambrogio, F.}, \bibinfo{year}{2015}.
\newblock \bibinfo{title}{Optimal transport for applied mathematicians}.
\newblock \bibinfo{publisher}{Birkh{\"a}user}.
%Type = Article
\bibitem[{Schl{\"a}pfer et~al.(2021)Schl{\"a}pfer, Dong, O'Keeffe, Santi,
  Szell, Salat, Anklesaria, Vazifeh, Ratti and West}]{schlapfer2021universal}
\bibinfo{author}{Schl{\"a}pfer, M.}, \bibinfo{author}{Dong, L.},
  \bibinfo{author}{O'Keeffe, K.}, \bibinfo{author}{Santi, P.},
  \bibinfo{author}{Szell, M.}, \bibinfo{author}{Salat, H.},
  \bibinfo{author}{Anklesaria, S.}, \bibinfo{author}{Vazifeh, M.},
  \bibinfo{author}{Ratti, C.}, \bibinfo{author}{West, G.B.},
  \bibinfo{year}{2021}.
\newblock \bibinfo{title}{The universal visitation law of human mobility}.
\newblock \bibinfo{journal}{Nature} \bibinfo{volume}{593},
  \bibinfo{pages}{522--527}.
%Type = Article
\bibitem[{Shang et~al.(2022)Shang, Yang, Yao, Tong, Yang and
  Mi}]{shang2022integrated}
\bibinfo{author}{Shang, P.}, \bibinfo{author}{Yang, L.}, \bibinfo{author}{Yao,
  Y.}, \bibinfo{author}{Tong, L.C.}, \bibinfo{author}{Yang, S.},
  \bibinfo{author}{Mi, X.}, \bibinfo{year}{2022}.
\newblock \bibinfo{title}{Integrated optimization model for hierarchical
  service network design and passenger assignment in an urban rail transit
  network: A {L}agrangian duality reformulation and an iterative layered
  optimization framework based on forward-passing and backpropagation}.
\newblock \bibinfo{journal}{Transportation Research Part C: Emerging
  Technologies} \bibinfo{volume}{144}, \bibinfo{pages}{103877}.
%Type = Article
\bibitem[{Simini et~al.(2012)Simini, Gonz{\'a}lez, Maritan and
  Barab{\'a}si}]{simini2012universal}
\bibinfo{author}{Simini, F.}, \bibinfo{author}{Gonz{\'a}lez, M.C.},
  \bibinfo{author}{Maritan, A.}, \bibinfo{author}{Barab{\'a}si, A.L.},
  \bibinfo{year}{2012}.
\newblock \bibinfo{title}{A universal model for mobility and migration
  patterns}.
\newblock \bibinfo{journal}{Nature} \bibinfo{volume}{484},
  \bibinfo{pages}{96--100}.
%Type = Article
\bibitem[{Song et~al.(2010)Song, Qu, Blumm and Barab{\'a}si}]{song2010limits}
\bibinfo{author}{Song, C.}, \bibinfo{author}{Qu, Z.}, \bibinfo{author}{Blumm,
  N.}, \bibinfo{author}{Barab{\'a}si, A.L.}, \bibinfo{year}{2010}.
\newblock \bibinfo{title}{Limits of predictability in human mobility}.
\newblock \bibinfo{journal}{Science} \bibinfo{volume}{327},
  \bibinfo{pages}{1018--1021}.
%Type = Article
\bibitem[{Taha and Hanbury(2015)}]{taha2015efficient}
\bibinfo{author}{Taha, A.A.}, \bibinfo{author}{Hanbury, A.},
  \bibinfo{year}{2015}.
\newblock \bibinfo{title}{An efficient algorithm for calculating the exact
  {H}ausdorff distance}.
\newblock \bibinfo{journal}{IEEE Transactions on Pattern Analysis and Machine
  Intelligence} \bibinfo{volume}{37}, \bibinfo{pages}{2153--2163}.
%Type = Article
\bibitem[{Tong et~al.(2015)Tong, Zhou and Miller}]{tong2015transportation}
\bibinfo{author}{Tong, L.}, \bibinfo{author}{Zhou, X.},
  \bibinfo{author}{Miller, H.J.}, \bibinfo{year}{2015}.
\newblock \bibinfo{title}{Transportation network design for maximizing
  space--time accessibility}.
\newblock \bibinfo{journal}{Transportation Research Part B: Methodological}
  \bibinfo{volume}{81}, \bibinfo{pages}{555--576}.
%Type = Misc
\bibitem[{{Urban Redevelopment Authority Singapore}(2019)}]{urapreviousplan}
\bibinfo{author}{{Urban Redevelopment Authority Singapore}},
  \bibinfo{year}{2019}.
\newblock \bibinfo{title}{{Singapore Master Plan: Previous Master Plans}}.
\newblock \bibinfo{note}{Available at:
  \url{https://www.ura.gov.sg/Corporate/Planning/Master-Plan/Previous-Master-Plans}}.
%Type = Book
\bibitem[{Vuchic(2017)}]{vuchic2017urban}
\bibinfo{author}{Vuchic, V.R.}, \bibinfo{year}{2017}.
\newblock \bibinfo{title}{Urban transit: Operations, planning, and economics}.
\newblock \bibinfo{publisher}{John Wiley \& Sons}.
%Type = Article
\bibitem[{Wang and Lo(2010)}]{wang2010global}
\bibinfo{author}{Wang, D.Z.}, \bibinfo{author}{Lo, H.K.}, \bibinfo{year}{2010}.
\newblock \bibinfo{title}{Global optimum of the linearized network design
  problem with equilibrium flows}.
\newblock \bibinfo{journal}{Transportation Research Part B: Methodological}
  \bibinfo{volume}{44}, \bibinfo{pages}{482--492}.
%Type = Article
\bibitem[{Wu et~al.(2024)Wu, Zuo, Zhou, Wan, Zhao and Yang}]{wu2024multi}
\bibinfo{author}{Wu, B.}, \bibinfo{author}{Zuo, X.}, \bibinfo{author}{Zhou,
  M.}, \bibinfo{author}{Wan, X.}, \bibinfo{author}{Zhao, X.},
  \bibinfo{author}{Yang, S.}, \bibinfo{year}{2024}.
\newblock \bibinfo{title}{A multi-objective ant colony system-based approach to
  transit route network adjustment}.
\newblock \bibinfo{journal}{IEEE Transactions on Intelligent Transportation
  Systems} \bibinfo{volume}{25}, \bibinfo{pages}{7878 -- 7892}.
%Type = Article
\bibitem[{Yu et~al.(2023)Yu, Chen, Liu and Zhu}]{yu2023urban}
\bibinfo{author}{Yu, X.}, \bibinfo{author}{Chen, Z.}, \bibinfo{author}{Liu,
  F.}, \bibinfo{author}{Zhu, H.}, \bibinfo{year}{2023}.
\newblock \bibinfo{title}{How urban metro networks grow: From a complex network
  perspective}.
\newblock \bibinfo{journal}{Tunnelling and Underground Space Technology}
  \bibinfo{volume}{131}, \bibinfo{pages}{104841}.
%Type = Inproceedings
\bibitem[{Zhou et~al.(2023)Zhou, Tan, Yean, Schl\"apfer and Lee}]{10508873}
\bibinfo{author}{Zhou, J.}, \bibinfo{author}{Tan, S.X.}, \bibinfo{author}{Yean,
  S.}, \bibinfo{author}{Schl\"apfer, M.}, \bibinfo{author}{Lee, B.S.},
  \bibinfo{year}{2023}.
\newblock \bibinfo{title}{Mobility patterns before, during and after the
  {COVID-19} pandemic in {S}ingapore}, in: \bibinfo{booktitle}{11th
  International Conference on Traffic and Logistic Engineering (ICTLE)}, pp.
  \bibinfo{pages}{129--134}.

\end{thebibliography}

\end{document}